\documentclass[manuscript]{aastex}
\usepackage{graphicx,epsfig,float,rotating,multirow}
\usepackage[caption=false]{subfig}

\shorttitle{Study of Three Early Type Semi-detached Systems: XZ Aql, UX Her and AT Peg}
\shortauthors{Zola et al.}

\begin{document}

\def\astrobj#1{#1}
\title{Photometric, Spectroscopic and Orbital Period Study of\\
Three Early Type Semi-detached Systems:\\
 \astrobj{XZ~Aql}, \astrobj{UX~Her} and \astrobj{AT~Peg}}


\author{S. Zola\altaffilmark{1,2}}
\affil{Astronomical Observatory, Jagiellonian University}
\author{\"O. Ba\c{s}t\"urk\altaffilmark{3}}
\affil{Ankara University, Faculty of Science}
\author{A. Liakos\altaffilmark{4}}
\affil{National Observatory of Athens}
\author{K. Gazeas\altaffilmark{5}}
\affil{National \& Kapodistrian University of Athens}
\author{H.~V. \c{S}enavc{\i}\altaffilmark{3}}
\affil{Ankara University, Faculty of Science}
\author{R.~H. Nelson\altaffilmark{6,7}}
\affil{Guest investigator, Dominion Astrophysical Observatory}
\author{\.{I}. \"Ozavc{\i}\altaffilmark{3}}
\affil{Ankara University, Faculty of Science}
\author{B. Zakrzewski\altaffilmark{2}}
\affil{Mt Suhora Observatory, Pedagogical University}
\and
\author{M. Y{\i}lmaz\altaffilmark{3}}
\affil{Ankara University, Faculty of Science}

\altaffiltext{1}{Astronomical Observatory, Jagiellonian University, ul. Orla 171, PL-30-244 Krakow, Poland}
\altaffiltext{2}{Mt Suhora Observatory, Pedagogical University, ul. Podchorazych 2, PL-30-084 Krakow, Poland}
\altaffiltext{3}{Ankara University, Faculty of Science, Dept. of Astronomy and Space Sciences, Tando\u{g}an, TR-06100, Ankara, Turkey}
\altaffiltext{4}{Institute for Astronomy, Astrophysics, Space Applications and Remote Sensing, National Observatory of Athens, Penteli, Athens, Greece}
\altaffiltext{5}{Department of Astrophysics, Astronomy and Mechanics, National \& Kapodistrian University of Athens, Zografos, Athens, Greece}
\altaffiltext{6}{1393 Garvin Street, Prince George, BC V2M 3Z1, Canada}
\altaffiltext{7}{Dominion Astrophysical Observatory, Herzberg Institute of Astrophysics, National Research Council of Canada}


\begin{abstract}
In this paper we present a combined photometric, spectroscopic
 and orbital period study of three early-type eclipsing binary systems: 
\astrobj{XZ~Aql}, \astrobj{UX~Her}, and \astrobj{AT~Peg}. 
As a result, we have derived the absolute parameters of their components and, 
on that basis, we discuss their evolutionary states. Furthermore, 
we compare their parameters with those of other binary systems and with the 
theoretical models. An analysis of all available up-to-date times of 
minima indicated that all three systems 
studied here show cyclic orbital changes; their origin is discussed 
in detail. Finally, we performed a frequency analysis for possible 
pulsational behavior and as a result we suggest that \astrobj{XZ~Aql} 
hosts a $\delta$ Scuti component.
\end{abstract}





\section{Introduction}
Classification of eclipsing binaries is performed according to the physical and 
evolutionary characteristics of their components, in addition to the shapes of 
their light curves. The degree to which their inner critical equipotential 
surfaces (Roche lobes) have been filled is a critical parameter for their 
classification, which helps in the understanding of their physical nature. 
Semi-detached binaries constitute an important class of objects, with one 
component filling its Roche lobe. The shape of a typical semi-detached binary 
light curve is an Algol-type light variation. These objects are important in 
understanding this stage of evolution when the mass transfer starts to take 
place and alters the evolution of the components as single stars. Whether or 
not one or both of the components are in contact with their Roche lobes or 
very close to filling them is very important in understanding the evolution 
of interacting binary systems. 
In order to achieve this goal, light and radial velocity 
observations are analyzed to determine their absolute parameters (temperatures, 
masses, radii, and surface potentials in particular). If the mass transfer has 
started in these systems at some point during their evolution, it also 
manifests itself as orbital period changes because a transfer of mass 
would alter a system's moment of inertia.\\

In this study, we present the results of light curve and period change 
analyses of three Algol-type binary systems: \astrobj{XZ~Aql}, 
\astrobj{UX~Her}, and \astrobj{AT~Peg}. We derive their absolute parameters 
from the analysis of their light and radial velocity curves that we obtained 
at two different observatories. We analyze the differences between the 
observed (O) and the calculated (C) eclipse timings, occuring 
due to the changes in their orbital period, by constructing O-C diagrams.
We also perform a frequency analysis to investigate the potential pulsation 
signals in the data of the studied systems. 
Finally, we discuss the evolutionary states of the components of these 
systems on Hertzsprung-Russell, Mass-Luminosity and Mass-Radius diagrams 
(hereafter HRD, MLD, and MRD, respectively). Such thorough analyses 
for these systems were being performed for the first time within this study.    

\section[]{Systems}
\subsection{\astrobj{XZ Aql}}
\astrobj{XZ Aql} (HD 193740, BD-07$^{\circ}$ 5271, GSC 5174-108, SAO 144345)
is an Algol type eclipsing binary. It was discovered by \citet{cannon22}.
The first detailed description of the light curve (without a plot) was 
presented by \citet{witkowski25}. \citet{erleksova59} was the first 
(and, so far the only one) to present the graphical light curve and give the 
first discussion of the O-C diagram. She proposed two alternative models 
of the period variation: first, as an abrupt change between 
JD 2430000 and 2433000, and second as an increase at a constant rate. 
\citet{pokzla76} discussed the same subject, using a larger
number of eclipse timings. \citet{woodforbes63} and \citet{samolyk96} 
noted a quadratic term in the ephemeris. A detailed discussion of shape of 
O-C diagram was given by \cite{soydugan06}. They modeled the variation 
with a sinusoidal superimposed on a secular parabolic change. They attributed 
the sinusoid variation to the light time effect caused by an unseen third body, 
and the secular term to a conservative mass transfer from the less massive 
component to the more massive one, which had already been noted by previous 
studies. The orbital period of the hypothetical third body was 
P$_{3}$ = 36.7$\pm$0.6 yr. The spectral type of XZ Aql is A2, 
found by \citet{cannon22} when the star was discovered. 
The mass ratio based on radial velocity observations has not been 
published until now, nor has any light curve solution.

\subsection{\astrobj{UX Her}}
\astrobj{UX Her} (HD 163175, BD+16$^{\circ}$ 3311, HIP 87643, SAO 103195,  
GSC 01557-01268) system was discovered by Cannon \citep{pickering08}, 
who found its spectral type as B9 or A.
\citet{zinner13} confirmed the discovery and determined the correct period
of the system. Later, \citet{tsesevich44}, \citet{kaho52}, 
\citet{kurzemnietse52}, \citet{ashbrook52}, \citet{tsesevich54}, 
\citet{koch62}, revised the elements. 
\citet{gordonkron63} and \citet{gordonkron65} published the first 
light curve solution based on the spectroscopic orbit determined by 
\citet{sanford37} and their own photometric observations. They 
proposed that the secondary was an evolved, low-mass, late-type star. 
\citet{hilletal75} estimated the primary component's spectral type as 
A0V - A3V for different orbital phases. 
Following the light curve studies of \citet{cester79}, \citet{mardirossian80} 
and \citet{giurmar81}, \citet{lazaro97} analyzed the first 
infrared light curves of the system in J, H and K bands, together with 
published B and V band light curves of \citet{gordonkron65}.
They determined the parameters of the system and found that none of the 
components of the binary filled their Roche lobes.  Although \astrobj{UX Her} 
is a short-period system, they assigned it to a category of 
slightly detached systems, most of which, as they pointed out, 
were long-period systems. \citet{gojko06} computed the mass ratio 
(q = m$_{2}$ / m$_{1}$ = 0.248) as the result of the  q-search method. 
When combined with the results of their B and V band light curve 
analysis, this mass ratio value suggests a semi-detached configuration 
of \astrobj{UX Her}.\\

\citet{kurzemnietse52}  noted for the first time that the period was variable.
\citet{tremko04} first published  a period-variation study of the system, 
which excluded a mass transfer between the components as the cause 
of the observed variations in the orbital period of the system, since 
neither of the stars filled their Roche lobes according to their interpretation.
They proposed that a low mass (0.3 $M_\odot$) unseen companion, bound to the 
system, was causing the period to change periodically. 

\subsection{\astrobj{AT Peg}}
Cannon \citep{canpick24} published the first spectroscopic observation of 
\astrobj{AT~Peg} (HD 210892, BD+07$^{\circ}$ 4824, SAO 127380, GSC 1137-185), 
and determined its spectral type as A0. 
The variability of the system was first announced 
by \citet{schneller31}, who identified it as an Algol-type binary. 
\citet{guthnick31}, \citet{rugemer34}, \citet{lass35}, \citet{criswal63}, 
\citet{woodforbes63}, \citet{oburka64a}, \citet{oburka64b}, \citet{oburka65} 
and \citet{criswal65}  published their photometric observations and revised 
the light elements. \citet{hillbarnes72} published the orbital elements of 
the system based on the first detailed spectroscopic observations 
and determined the spectral type as A7 V. 
They found the mass ratio ($m_{1}$ / $m_{2}$) to be 
2.4 and the orbital inclination 76$^{\circ}$.7. \citet{gulmenetal93} found 
that AT Peg was a semi-detached binary with a later type subgiant secondary 
component filling its Roche lobe. 
\citet{maxtedetal94} obtained spectra of the system and determined 
a spectroscopic mass ratio of 2.115 ($m_{1}$ / $m_{2}$), smaller than the 
one determined by \citet{hillbarnes72} from photographic plates.
They determined absolute parameters of the system using the photometric 
observations of \citet{criswal63}. They also classified
the spectral type of the primary component (A4 V) from a combined spectrum of
their data, which was consistent with the temperature estimates
obatined from Str\"omgren photometry by \citet{hilditch75}.
The configuration of the system as a semi-detached eclipsing binary has 
been widely accepted since this study and that of \citet{giuricin81}.\\ 

\citet{savedoff51} was the first to notice that the orbital period 
was variable. Although the secular change in the eclipse timing variations 
has been noted by recent studies \citep{margrave81,guduretal87,liakos12a,
hanna12}, \citet{liakos12a} noticed also the discrepancy
between the configuration 
and the direction of the mass transfer. They suggested either mass loss via 
stellar winds or system angular momentum loss via magnetic braking 
as possible explanations. Periodic variations in the O-C diagram 
have been also noted and attributed to unseen third bodies with parameters 
differing from one study to another \citep{borhege95,borhege96,liakos12a},
or to magnetic activity with single \citep{sarna97}, or to multiple cycles 
\citep{hanna12}. However there is no firm evidence for strong magnetic activity 
other than enhanced X-ray emission \citep{hanna12}.

\section{Observations}

%% In a manner similar to \objectname authors can provide links to dataset
%% hosted at participating data centers via the \dataset{} command.  The
%% second curly bracket argument is printed in the text while the first
%% parentheses argument serves as the valid data set identifier.  Large
%% lists of data set are best provided in a table (see Table 3 for an example).
%% Valid data set identifiers should be obtained from the data center that
%% is currently hosting the data.
%%
%% Note that AASTeX interprets everything between the curly braces in the 
%% macro as regular text, so any special characters, e.g. "#" or "_," must be 
%% preceded by a backslash. Otherwise, you will get a LaTeX error when you 
%% compile your manuscript.  Special characters do not 
%% need to be escaped in the optional, square-bracket argument.
Between 2012 and 2013 we performed photometric observations of \astrobj{XZ~Aql},
 \astrobj{UX~Her}, and \astrobj{AT~Peg}, using the 40-cm 
Cassegrain telescope (f/5 using a focal reducer) of the 
Gerostathopoulion Observatory of the University of Athens (UoA Observatory). 
This setup results in a field-of-view (FOV) of 17$\times$26 arcmin. 
The telescope was equipped with with an SBIG ST-10 XME CCD detector and a set 
of $BVRI$ filters (Bessell specifications) in order to perform 
multi-band photometry. We present a log of our observations in 
Table~\ref{table1}.
\\
The ephemerides of all systems (see Table~\ref{table2}) were calculated using 
the least squares method 
on the minima timings derived from our observations 
and the most recent ones taken from the literature. The photometric data 
sets were reduced with dark and flat frames which were gathered 
before or after each observing run, while image reduction as well as 
differential aperture photometry were performed using either 
C-munipack \citep{hroch98, motl04} or AIP4WIN \citep{berry00} software packages.
 Comparison and check stars are also listed in Table~\ref{table2} 
together with their properties.\\

%------------------------------------------------------------Table 1
\begin{table*}
\scriptsize
\begin{center}
\caption{Log of photometric observations}
\label{table1}
\medskip
\begin{tabular}{lccccccccc}
\hline
\hline
System & Dates of observations & \multicolumn{4}{c}{Number of data points} & \multicolumn{4}{c}{Mean errors} \\
  &  & B & V & R & I & $\sigma_{B}$(mag) & $\sigma_{V}$(mag) & $\sigma_{R}$(mag) & $\sigma_{I}$(mag) \\
\hline
 \astrobj{XZ~Aql}   & 26 nights in  & 2282 & 2351 & 2405 & 2379 & 0.0091 & 0.0090 & 0.0086 & 0.0075 \\
                    & Jun-Aug 2013  &   &   &   &   &   &   &   &   \\
 \astrobj{UX~Her}   & 10 nights in  & 2056 & 1986 & 1984 & 2028 & 0.0035 & 0.0032 & 0.0031 & 0.0034 \\
                    & Jun 2012      &   &   &   &   &   &   &   &   \\
 \astrobj{AT~Peg}   & 6 nights & 895 & 895 & 895 & 895 & 0.0047 & 0.0054 & 0.0050 & 0.0054 \\
& in Aug 2012 & & & & & & & & \\

\hline
\end{tabular}
\end{center}
\end{table*}
%-------------------------------------------------------

%------------------------------------------------------------Table 2
\begin{table*}
\scriptsize
\begin{center}
\caption{Ephemerides, magnitudes of the system, and the  comparison 
and check stars}
\label{table2}
\medskip
\begin{tabular}{lcccccccccc}
\hline
\hline
                   & Epoch (T$_{0}$) & Period   & m$_{v}$ & (B-V) & 
\multicolumn{3}{c}{Comparison} & \multicolumn{3}{c}{Check} \\
System       & (HJD+2400000) &  (days) & (mag) & (mag) & 
GSC ID & m$_{v}$ & (B-V) & GSC ID & m$_{v}$ & (B-V)\\
\hline
\astrobj{XZ~Aql}   & 52501.0881 & 2.139207 & 10.18 & 0.25 & 
5175-0726 & 10.48 & 0.96 & 5174-0186 & 10.87 & 0.73 \\
\astrobj{UX~Her}   & 52501.5262 & 1.548869 &  8.97 & 0.15 & 
1557-1029 & 9.21 & 0.89 & 1557-1196 & 9.73 & 0.99 \\
\astrobj{AT~Peg}   & 52500.9285 & 1.146065 &  9.02 & 0.19 & 
1137-0492 & 9.78 & 0.77 & 1137-0134 & 10.58 & 0.53 \\
\hline
\end{tabular}
\end{center}
\end{table*}
%-------------------------------------------------------

We acquired radial velocity measurements of 
our targets at the Dominion Astrophysical Observatory (DAO) in Victoria, 
British Columbia, Canada using the Cassegrain spectrograph 
attached on the 1.85 m Plaskett telescope. A grating (\#21181) with 
1800 lines/mm, blazed at 5000 \AA~giving a reciprocal linear dispersion 
of 10 \AA/mm in the first order and covering 
a wavelength region from 5000 to 5260 \AA~approximately was used. 
A log of all spectroscopic observations is presented in Table~\ref{table3}. 
We used 'RaVeRe' software \citep{nelson2010} for cosmic hit removal, median 
background fitting and subtraction for each column, aperture summation, 
continuum normalization, wavelength calibration and linearization 
using the Fe-Ar spectra as wavelength standards. Finally, the radial velocities 
were extracted from the spectra by using the Rucinski broadening functions
\citep{rucinski04, nelson2010b, nelson2014} as implemented in the 
software 'Broad' \citep{nelson2014}. In this process, 
IAU standard radial velocity stars, of spectral types F and G, 
were used to solve directly for the functions that account for the 
broadening and Doppler shifts of the spectra.  
In practice, a number of standard stars were used individually, 
and the mean values of the resulting radial velocities were adopted. 
In order to investigate the effect of the differing spectral types 
of the standard stars on the resulting radial velocities, 
we split the standard stars into two groups: one with F-types, and the other, 
G-types. Looking at the mean radial velocities derived from each group, 
we noted negligible differences between corresponding values--certainly 
less than the estimated errors.  Therefore we did not deem it necessary 
to use different standard stars for primary and secondary spectra or to 
restrict the standard star choices in any other way.

\astrobj{AT Peg} has previous radial velocity measurements using the 
1.2 metre telescope at the DAO were reported in \citet{maxtedetal94}. 
Their determinations used an older technique involving a reticon detector 
and cross-correlations.  We made use of their measurements in our analysis 
but did not combine both radial velocity data sets together because the 
V$_{\gamma}$ values differ from one data set to another.

\section{Analysis}
\subsection{Light Curve Modeling}
In order to obtain initial parameters of the systems 
studied in this paper we used the Wilson-Devinney program augmented 
with the Monte Carlo search algorithm to ensure that the global minimum 
was found within the set ranges of free parameters. We followed the 
procedure outlined in \citet{zolaetal2014}, that is keeping the values 
constant for systems' mass ratios, as derived from sine fitting to the 
radial velocities. In the search for solutions, we adjusted inclination (i), 
temperature of the secondary (T$_{2}$), potentials ($\Omega_{1,2}$), 
luminosity of the primary star (L$_{1}$). Neither a spot nor a third 
light was required for the three targets analyzed in this work.
In addition to the mass ratios, we also kept the temperatures of the primary 
components (T$_{1}$) constant following from the spectral type of a 
system. In order to determine T$_{1}$, we used the spectral type 
versus temperature calibration published by \citet{Harmanec}. 
Furthermore, the albedo and gravity darkening coefficients were set 
at their theoretical values for either a radiative or a convective envelope. 
The limb darkening coefficients were built into the code and were chosen as a 
function of a star temperature and the wavelength of the observations from 
the tables by \citet{Claret2011}, \cite{Claret2012} and \cite{Claret2013} 
based on square root law. Due to the large number of observed single points, 
we calculated about 
150 mean points for every binary and in each filter. 
These were calculated in such a way that they evenly covered the observed light 
curve and using ephemerides listed in Table~\ref{table2}. 
This procedure speeded up computations and also provided 
error estimates for each mean point. 
The convergence was achieved for all three systems and the resulted 
configurations (not assumed a priori) were all semi-detached with 
the secondaries filling the Roche lobe. 
In the next step we used the resulting parameters as starting ones 
in the simultaneous solution of light and radial velocity curves using the 
latest version, 2015 release (WD2015), of the Wilson-Devinney code 
\citep{wildev71, wilson79, wilson90, vanhamme07, 
wilson08, wilson10, wilson12}. We made use of the Mode-5 for the configuration, 
the usual mode for Algol-type binaries with the secondary component filling its 
Roche lobe in parallel with our findings for the configurations 
in the previous step. The list of free parameters was similar to those of 
the Monte Carlo search, however, also the mass ratio was adjusted as well as 
further two parameters describing the orbit: the semi-major axis of binary 
system relative orbit and the systemic velocity.
We assumed that the orbits are circular due to strong tidal forces 
in case of semidetached systems.
These computations were done with all individual points and, setting the 
control parameter KSD=1, we made use of the program feature to let the program 
to compute the curve weights. Several iterations were required to derive 
the final, combined RV and LC solutions.

\begin{figure*}
\centering
\subfloat[]{
  \centering
  \includegraphics[width=84.00mm]{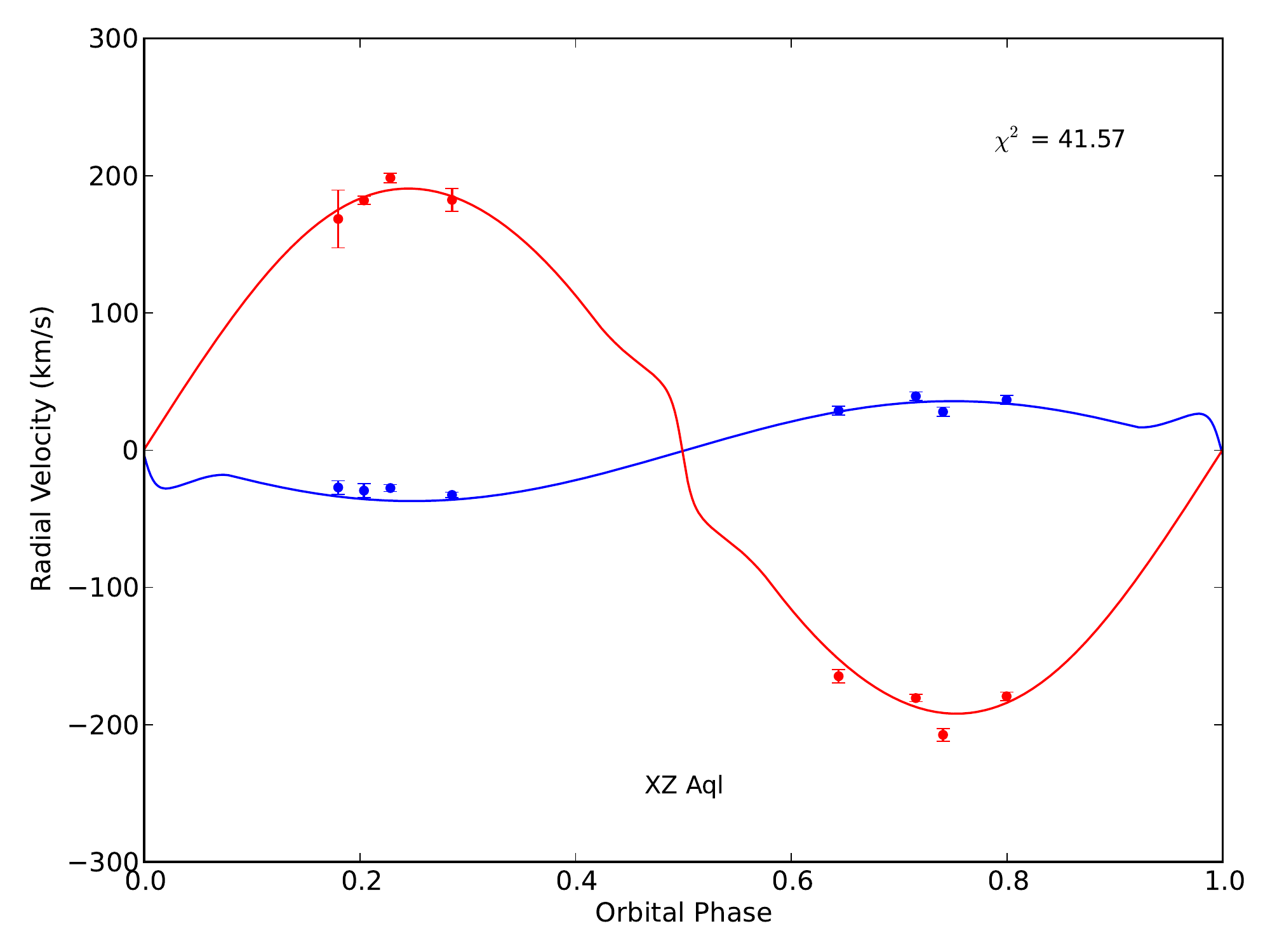}
  \label{fig_xzaql_rv}}
\subfloat[]{
  \centering
  \includegraphics[width=84.00mm]{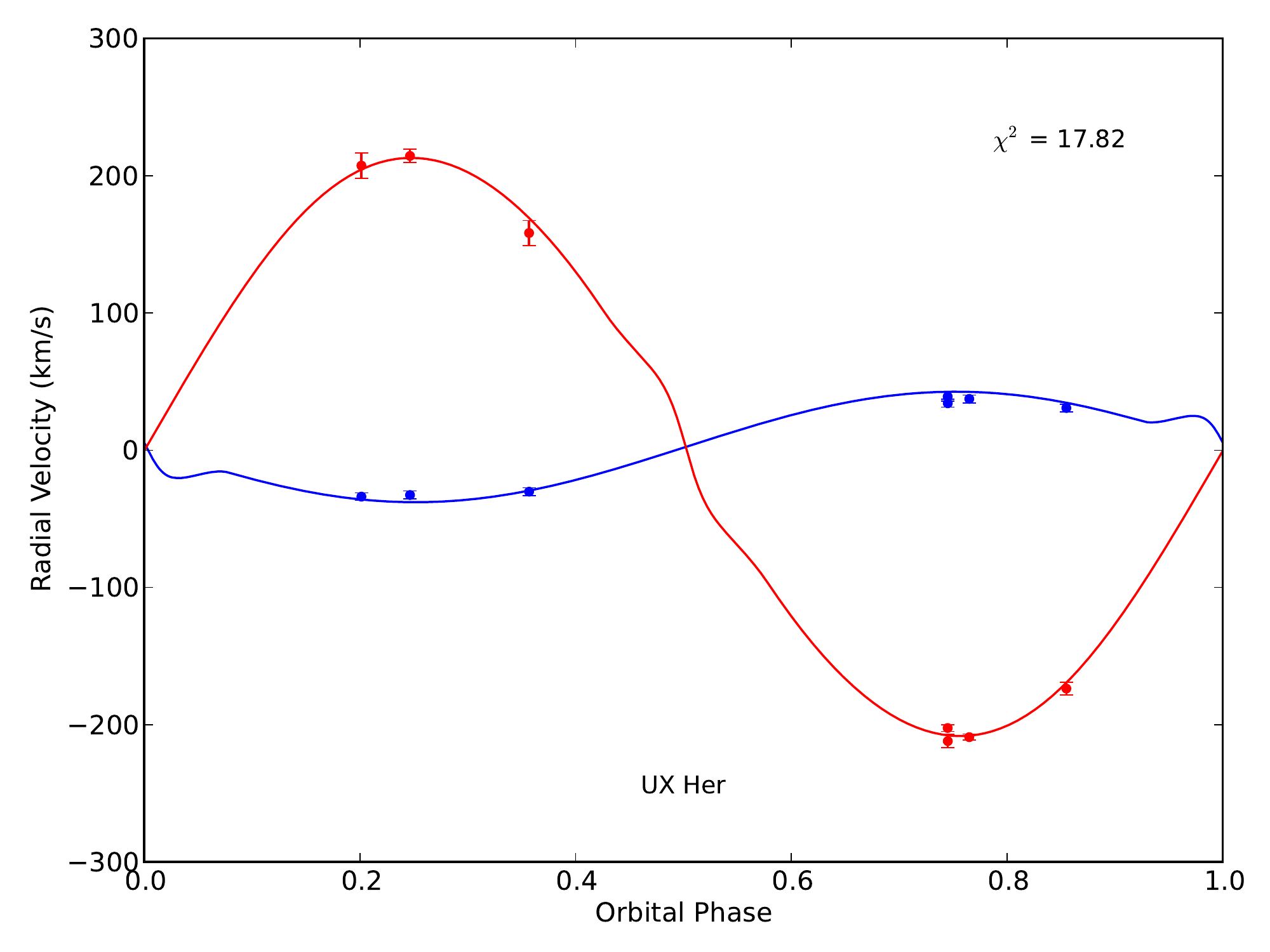}
  \label{fig_uxher_rv}}
\caption{Observed radial velocity curves of \astrobj{XZ~Aql} (a) 
and \astrobj{UX~Her} (b). The best fits to compute radial velocity 
amplitudes K1,K2 are also given in solid lines. Systemmic velocities have been 
subtracted from all the data.}
\label{fig_xzaql_uxher_rv}
\end{figure*}

\begin{figure*}
\centering
\subfloat[]{
  \centering
  \includegraphics[width=84.00mm]{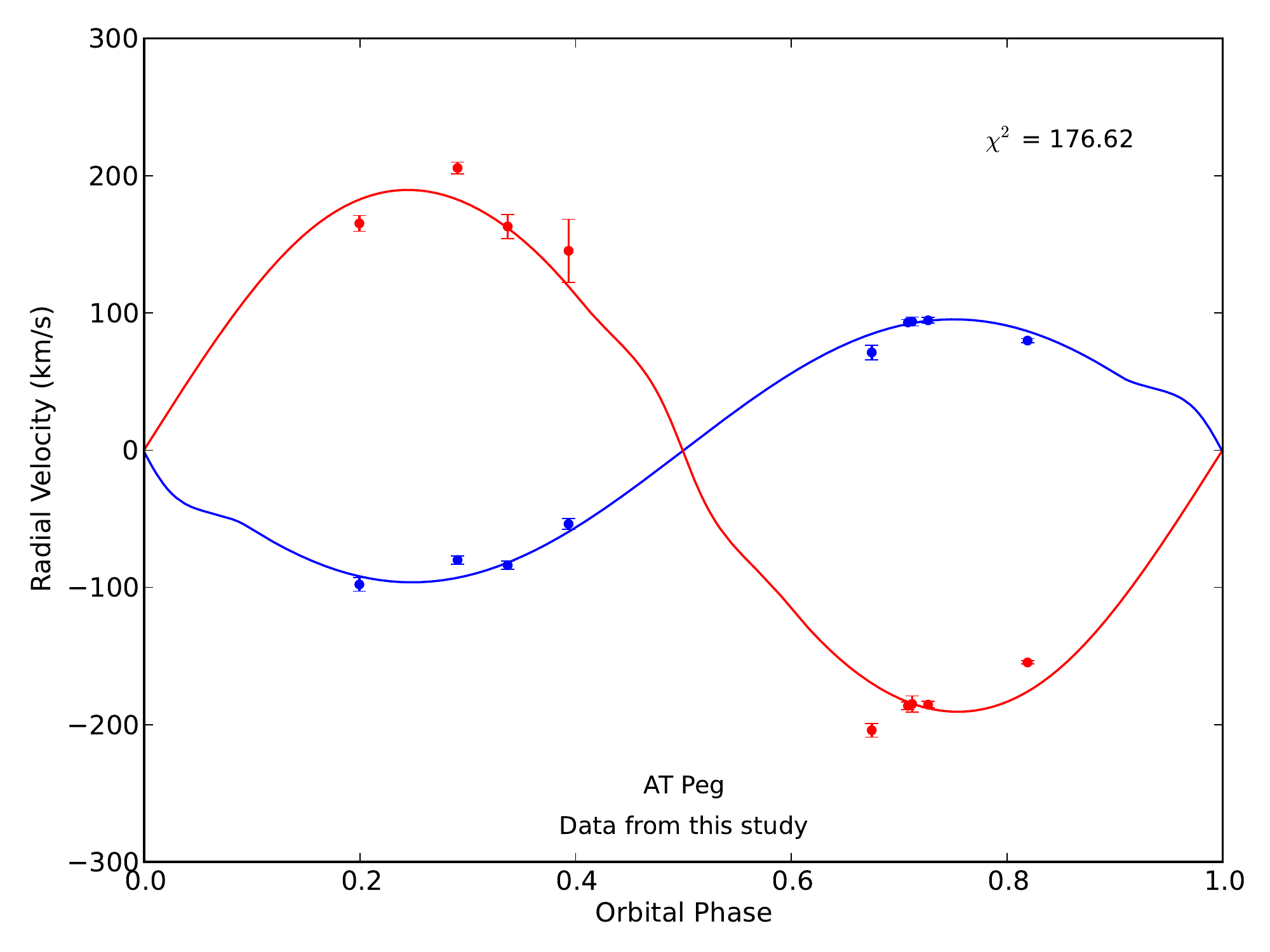}
  \label{fig_atpeg_bob_rv}}
\subfloat[]{
  \centering
  \includegraphics[width=84.00mm]{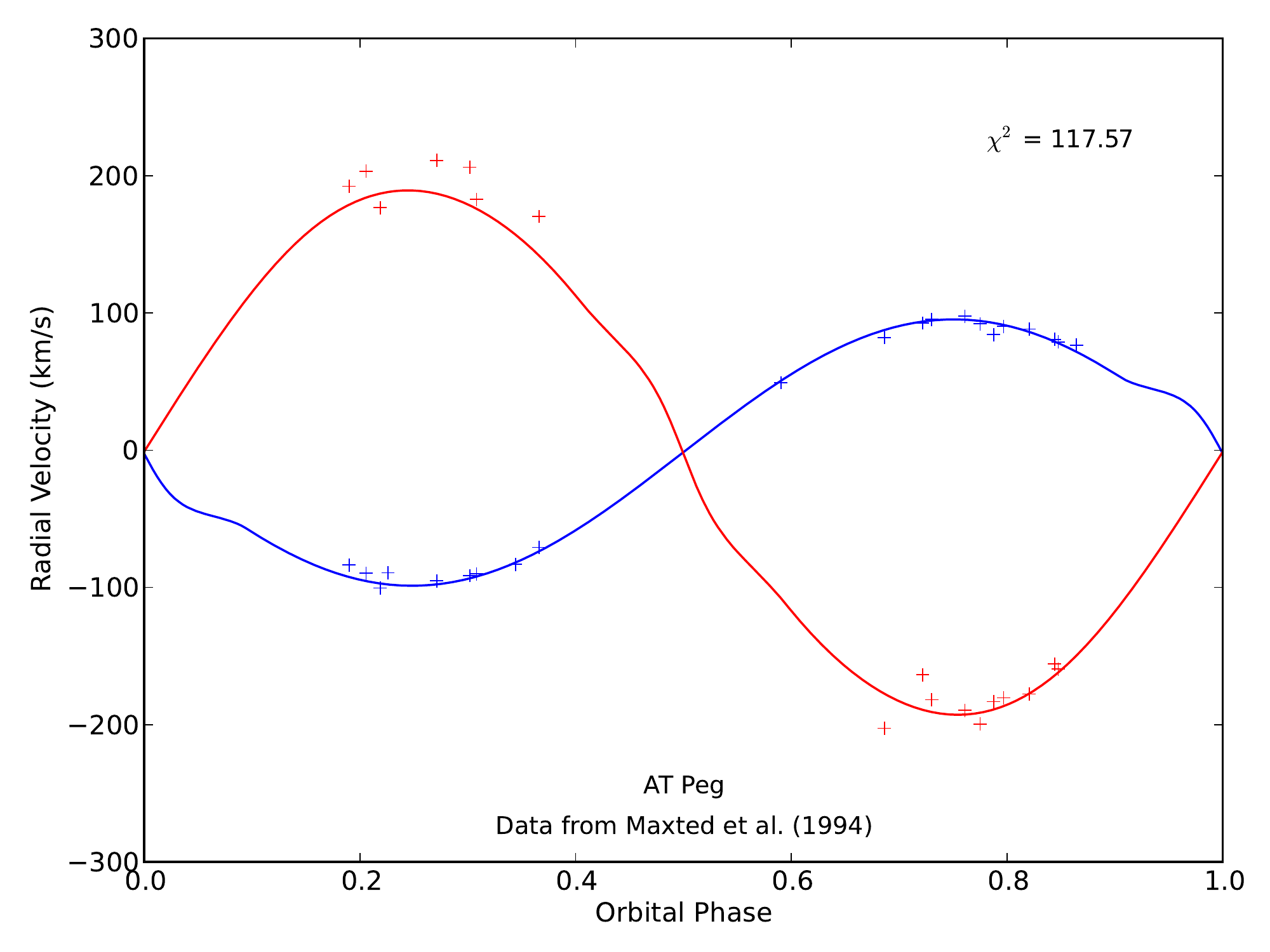}
  \label{fig_atpeg_maxted_rv}}
\caption{Same as in Fig.~\ref{fig_xzaql_uxher_rv}, 
but for two different data sets of \astrobj{AT~Peg}, 
our own measurements (a) and \citet{maxtedetal94}'s measurements 
(b).}
\label{fig_atpeg_rv}
\end{figure*}

\begin{figure*}
\centering
\includegraphics[width=170.00mm]{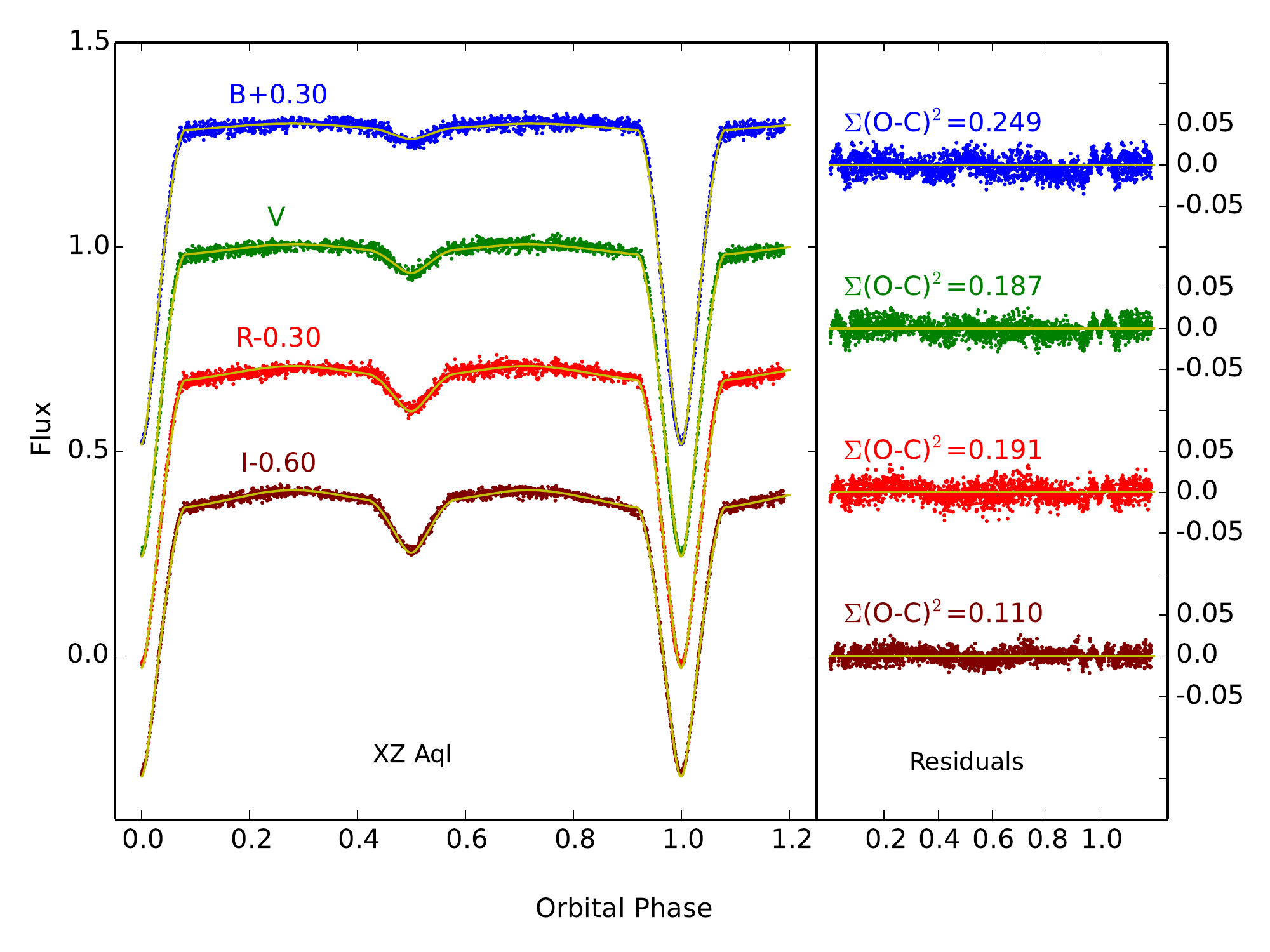}
\caption{Observed light curves of \astrobj{XZ~Aql}. Data in  $BVRI$ filters 
(from top to bottom) were shifted arbitrarily by the amounts shown in figures 
for clarity. Theoretical light curves corresponding to the best fits 
are ploted in solid lines. Residuals from the fits are given at the right of  
the light curves using the same symbols, and shifted also arbitrarily 
as the light curves.}
\label{fig_xzaql_lc}
\end{figure*}

\begin{figure*}
\centering
\includegraphics[width=170.00mm]{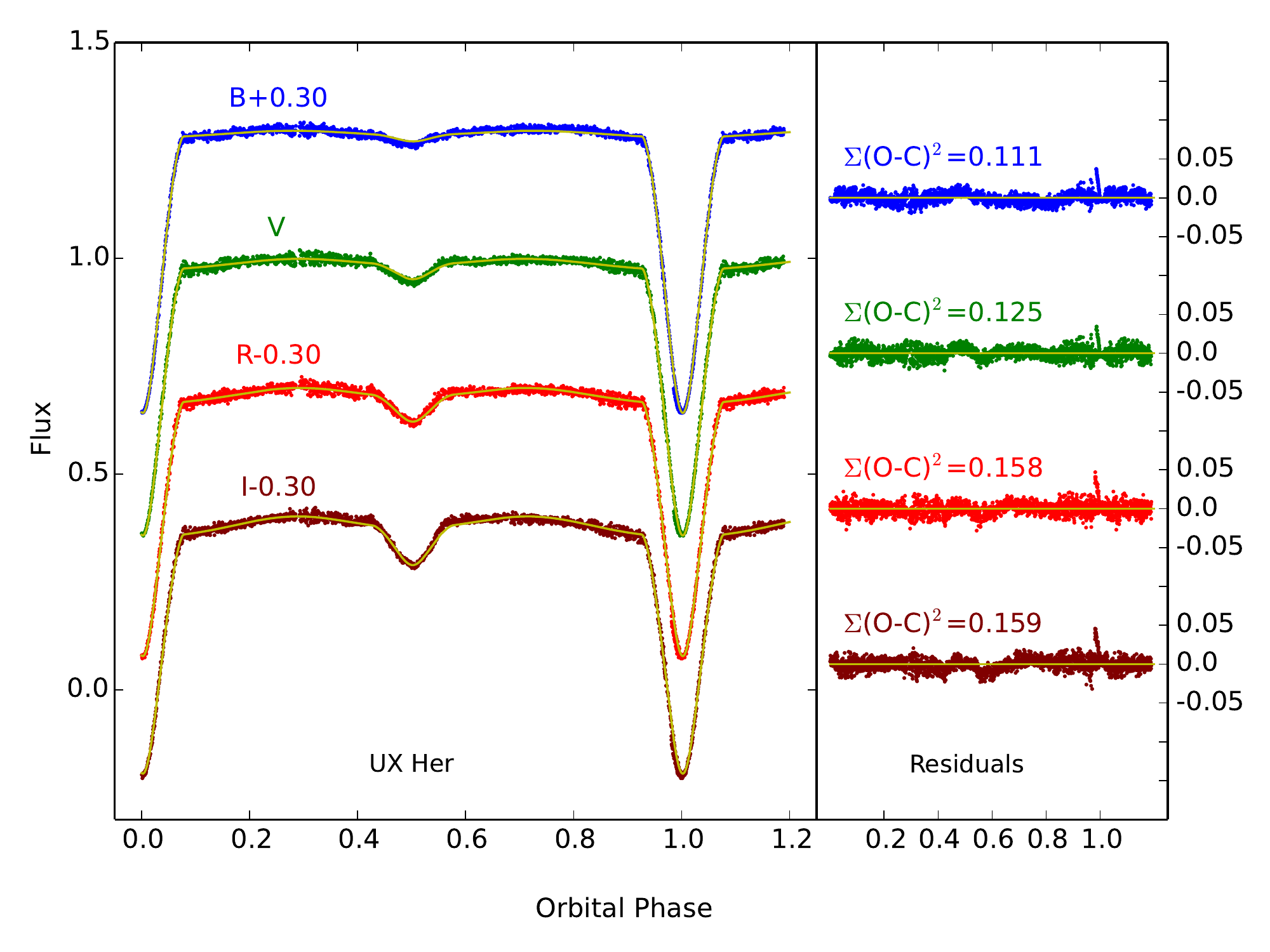}
\caption{Same as in Fig.~\ref{fig_xzaql_lc}, 
but for \astrobj{UX~Her}}
\label{fig_uxher_lc}
\end{figure*}

\begin{figure*}
\centering
\includegraphics[width=170.00mm]{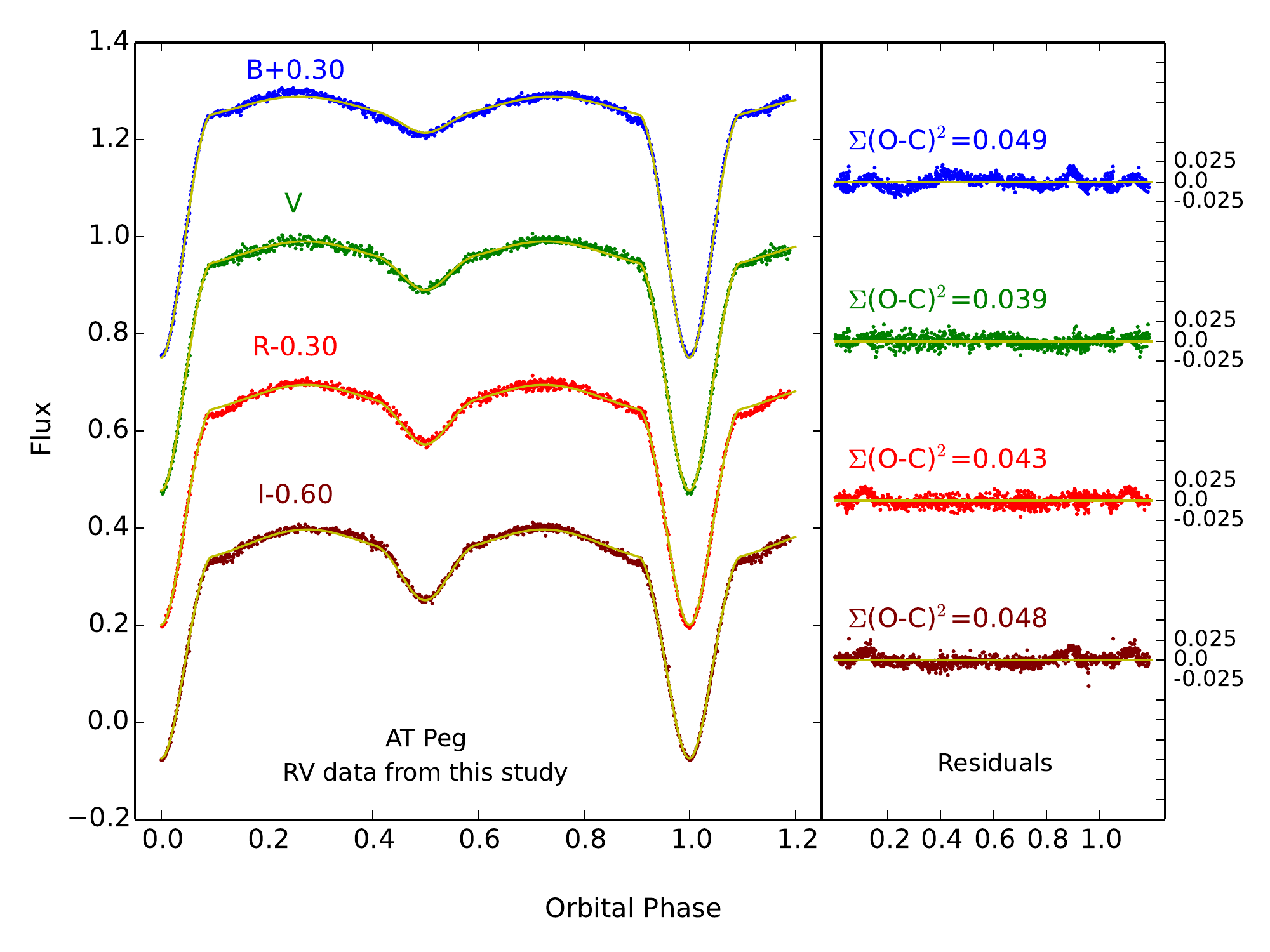}
\caption{Same as in Fig.~\ref{fig_xzaql_lc}, 
but for \astrobj{AT~Peg} using our own measurements in DAO.}
\label{fig_atpeg_bob_lc}
\end{figure*}

\begin{figure*}
\centering
\includegraphics[width=170.00mm]{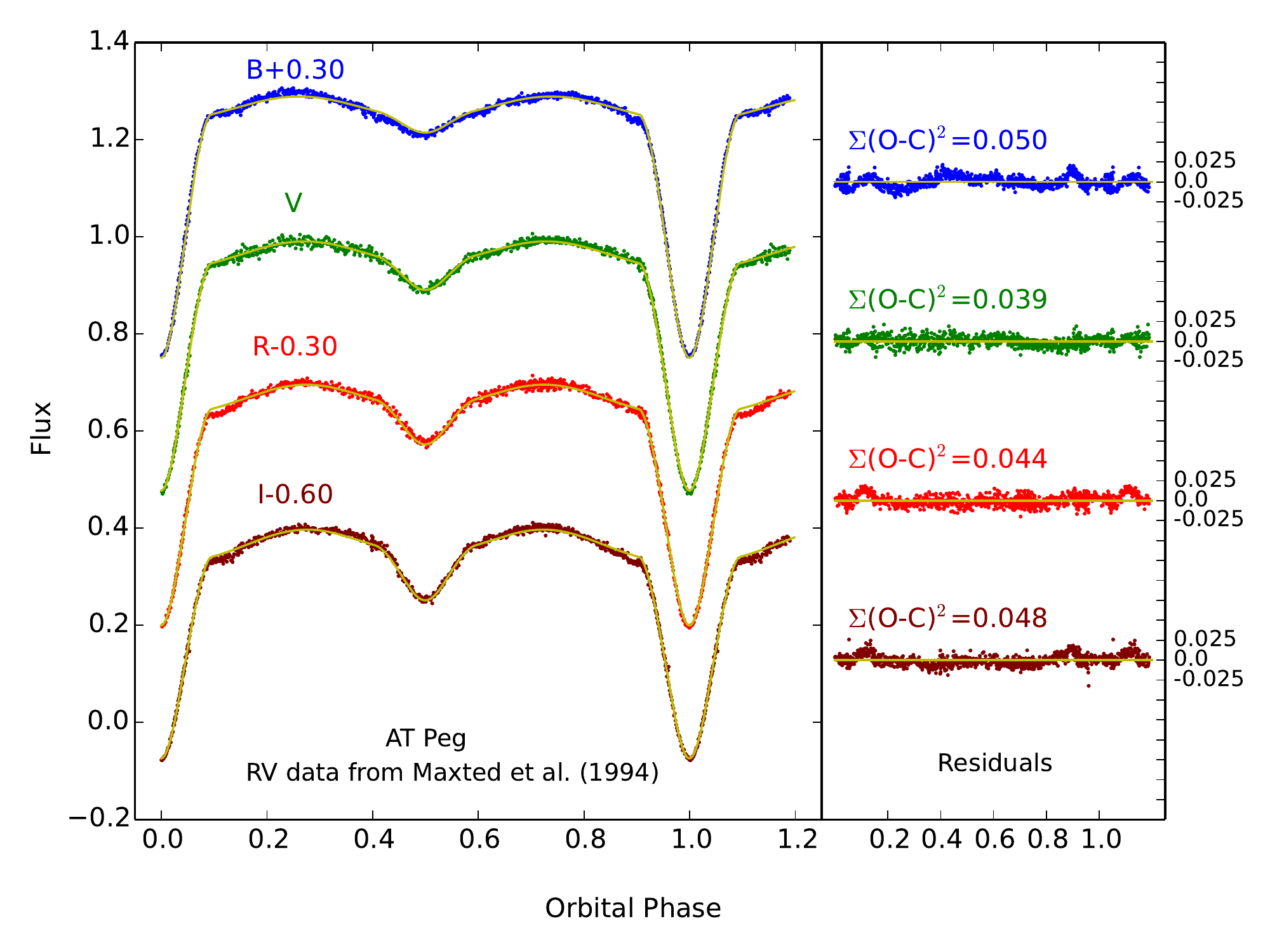}
\caption{Same as in Fig.~\ref{fig_xzaql_lc}, 
but for \astrobj{AT~Peg} using \citet{maxtedetal94}'s measurements.}
\label{fig_atpeg_maxted_lc}
\end{figure*}

For \astrobj{AT Peg}, we determined two separate solutions, 
one using RV data from \citet{maxtedetal94}, and the other, our own RV results.
All the radial velocity obervations and their fits obtained with the 
WD2015 code are given in Figs.~\ref{fig_xzaql_uxher_rv} - \ref{fig_atpeg_rv}.
The results of the light curve analyses, as derived in the second step, are 
listed in Table~\ref{table4} together with the formal error for 
each parameter in parenthesis as computed by the Wilson-Devinney code. 
The fits and their residuals are shown in Figs.~\ref{fig_xzaql_lc} - 
\ref{fig_atpeg_maxted_lc}. Finally, the computed absolute parameters are 
given in Table~\ref{table5}. \\

%----------------------------------------------------Figure XZ Aql - UX Her RV

%--------------------------------------------------- Figure AT Peg RV
%-----------------------------------------------------------------------------

%----------------------------------------------------Figure XZ Aql - UX Her LC

%--------------------------------------------------- Figure AT Peg LC
%------------------------------------------------------------Table 3
\begin{table}
\small
\begin{center}
\caption{Log of spectroscopic observations}
\label{table3}
\begin{tabular}{lcc}
\hline
\hline
System  & Dates of Obs. & \# of Data Pts. \\
\hline
\astrobj{XZ~Aql} & 26,27 Sep 2009; 1,2 Oct 2010; & 8 \\
 & 8,9,10,11 Sep 2013 & \\
\astrobj{UX~Her} & 22,23,26 Apr 2011; & 7 \\
 &  7,11,12 Sep 2011 & \\
\astrobj{AT~Peg} & 26,27,28 Sep 2009; 5 Oct 2010; & 9 \\
 & 4 Sep 2011; 8 Sep 2013 & \\
\hline
\end{tabular}
\end{center}
\end{table}
%-------------------------------------------------------

%------------------------------------------------------------Table 4
\begin{table*}
\small
\begin{center}
\caption{Results from the light curve analysis.*,**}
\label{table4}
\medskip
\begin{tabular}{l*{5}{c}}
\hline
\hline
Stellar Parameters & \astrobj{XZ~Aql} & \astrobj{UX~Her} & 
\multicolumn{2}{c}{\astrobj{AT Peg}} \\
& & & RV$_{1}$ & RV$_{2}$ \\
\hline
$i \ [deg] $   & 85.88(3) & 82.28(2) & 76.30(4) & 76.25(4) \\
$T_{1}$ [$K$] & 8770 & 8770 & 8360 & 8360 \\
$T_{2}$ [$K$] & 4744(5) & 4478(5) & 5057(6) & 5051(6) \\
$\Omega_{1}$ & 4.26(1) & 4.55(1) & 4.20(1) & 4.22(1) \\
$\Omega_{2}$ & 2.19(1) & 2.19(1) & 2.85(1) & 2.85(1) \\
V$_{\gamma} [km/s]$ & 17.6(9) & -67.6(11) & 10.0(19) & 4.1(6) \\
$q=m_{2}/m_{1}$ & 0.184(4) & 0.184(21) & 0.484(3) & 0.489(3) \\
\hline
Luminosities &  &  &  &  \\
\hline
$L_{1}$ [B] & 12.25(1) & 12.25(1) & 11.40(1) & 11.40(1) \\
$L_{1}$ [V] & 11.83(1) & 11.87(1) & 10.62(2) & 10.61(2) \\
$L_{1}$ [R] & 11.28(1) & 11.32(1) & 10.02(2) & 10.01(2) \\
$L_{1}$ [I] & 10.61(1) & 10.70(1) & 9.36(2) & 9.34(2)  \\
$L_{2}$ [B] & 0.30    & 0.23    & 0.82 & 0.83 \\
$L_{2}$ [V] & 0.70    & 0.56    & 1.56 & 1.56 \\
$L_{2}$ [R] & 1.21    & 1.04    & 2.17 & 2.18 \\
$L_{2}$ [I] & 1.77    & 1.63    & 2.82 & 2.83 \\
\hline
\end{tabular}
\end{center}
* Formal errors from the WD-code are in parantheses.\\
*** Column RV$_{1}$ shows the results from the analysis using our own 
RV measurements, and RV$_{2}$ shows the same but using RVs from 
\citet{maxtedetal94}.
\end{table*}
%-------------------------------------------------------
%------------------------------------------------------------Table 5
\begin{table*}
\small
\begin{center}
\caption{Absolute Parameters.*}
\label{table5}
\medskip
\begin{tabular}{l*{5}{c}}
\hline
\hline
Parameter & \astrobj{XZ~Aql}  & \astrobj{UX~Her} & 
\multicolumn{2}{c}{\astrobj{AT Peg}}\\
& & & RV$_{1}$ & RV$_{2}$ \\
\hline
$a [R_{\odot}]$ & 9.94(21) & 8.00(9) & 6.86(8) & 6.91(7)\\
$\mathcal{M}_{1} [M_{\odot}]$ & 2.42(14) & 2.42(12) & 2.22(8) & 2.26(7) \\
$\mathcal{M}_{2} [M_{\odot}]$ & 0.45(6) & 0.44(6) & 1.08(6) & 1.11(5) \\
$\mathcal{R}_{1} [R_{\odot}]$ & 2.45(7) & 1.84(5) & 1.86(3) & 1.87(3) \\
$\mathcal{R}_{2} [R_{\odot}]$ & 2.43(6) & 1.96(5) & 2.18(3) & 2.20(3) \\
M$_{bol,1}$ [mag] & 0.99(21) & 1.61(11) & 1.80(10) & 1.78(9) \\
M$_{bol,2}$ [mag] & 3.68(21) & 4.40(11) & 3.64(10) & 3.62(9) \\
log~g$_{1}$ [cgs] & 4.04(6) & 4.29(5) & 4.24(3) & 4.25(3) \\
log~g$_{2}$ [cgs] & 3.32(6) & 3.50(5) & 3.79(3) & 3.80(3) \\
\hline
\end{tabular}
\end{center}
* Column RV$_{1}$ shows the results from the analysis using our own 
RV measurements, and RV$_{2}$ shows the same but using RVs from 
\citet{maxtedetal94}.
\end{table*}

\subsection{Orbital Period Analysis}
\begin{figure*}
\centering
\subfloat[]{
  \centering
  \includegraphics[width=84.00mm]{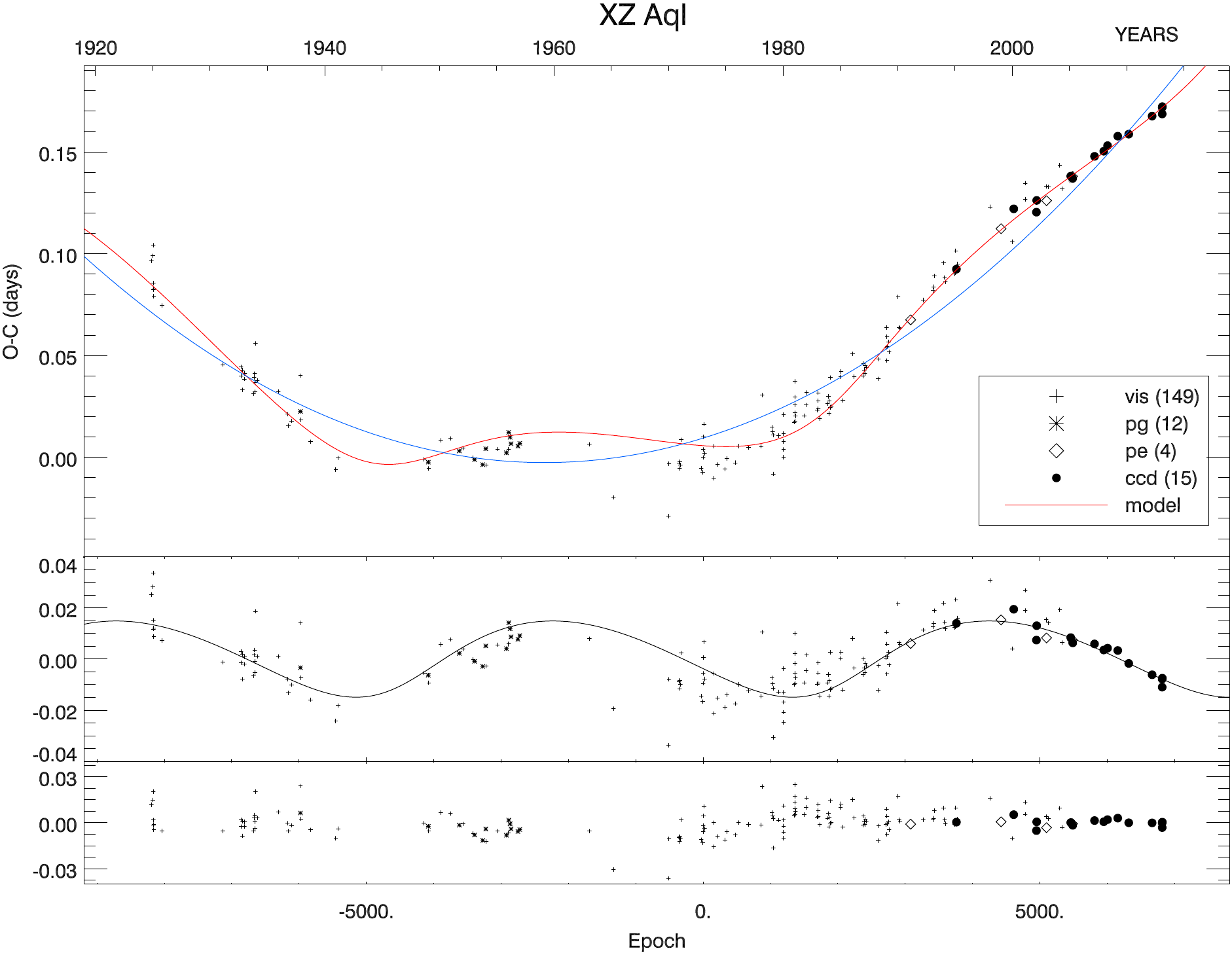}
  \label{fig_xzaql_oc}}
\subfloat[]{
  \centering
  \includegraphics[width=84.00mm]{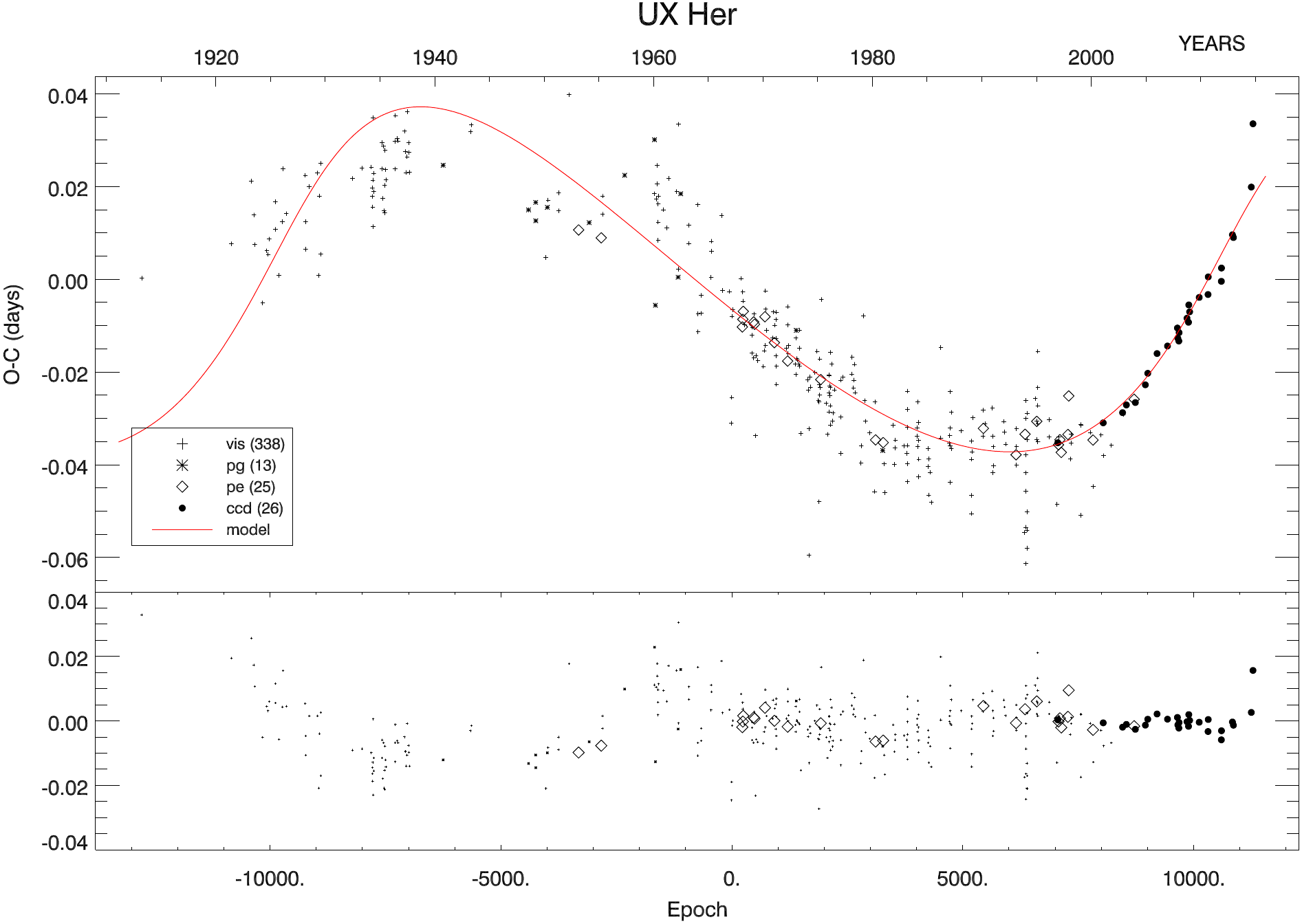}
  \label{fig_uxher_oc}}
\caption{The O-C diagrams for \astrobj{XZ~Aql} (a) and \astrobj{UX~Her} 
(b). The fits are plotted as solid lines 
(best fit models in blue and parabola fits superimposed by 
sinusoidal variations in red in the electronic version).
Residuals from the fits are given in the lower panels. 
Symbols have been defined in the legends.}. 
\label{fig_xzaql_uxher_oc}
\end{figure*}

\begin{figure*}
\centering
\includegraphics[width=85.00mm]{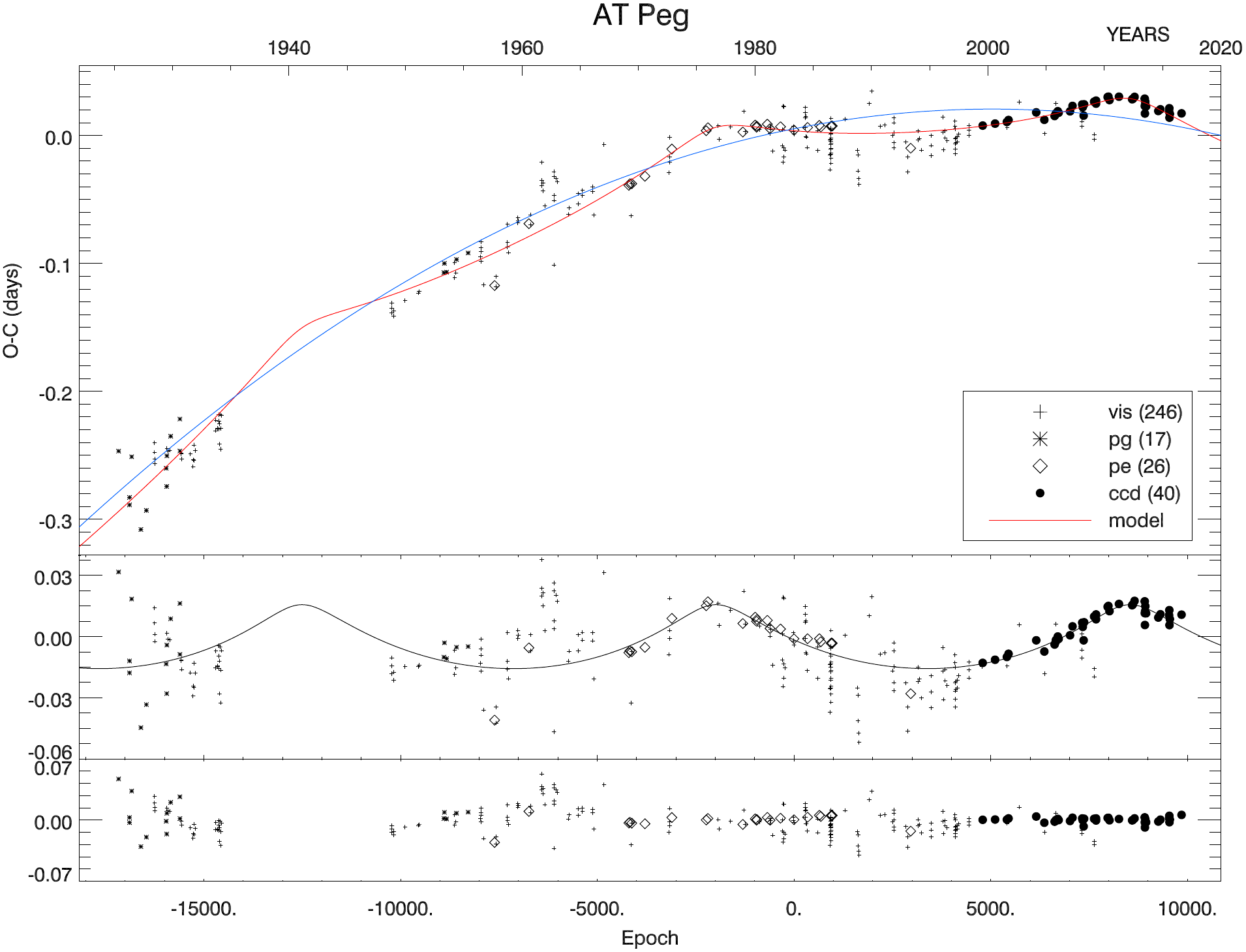}
\caption{Same as the Fig.~\ref{fig_xzaql_uxher_oc}, 
but for \astrobj{AT~Peg}}
\label{fig_atpeg_oc}
\end{figure*}

In order to better understand the nature of studied systems, we constructed 
the O-C diagrams (Figs.~\ref{fig_xzaql_uxher_oc} - \ref{fig_atpeg_oc}) 
by using all minima times available in the literature  
together with those derived from our own observations, 
and  weighting them according to the observation method 
(photographic: 0.3, visual: 0.5, photoelectric: 0.8, CCD: 1).  
We made use of the Kwee-van Woerden method 
\citep{kwee56} to derive 
the times of minimum light levels in our own light curves 
(Table~\ref{tablemins}).\\
%----------------------------------------------------- Table of Minima Times
\begin{table*}
\small
\begin{center}
\caption{Times of Minimum Light Levels Derived From Our Own Light Curves.}
\label{tablemins}
\medskip
\begin{tabular}{lcccc}
\hline
\hline
System & Time of Min. & Error & Filter & Type\\ 
 & (HJD +2400000)  & & & \\ 
\hline
XZ Aql & 56486.4291 & 0.0001 & BVRI & primary \\
 & 56487.4975 & 0.0004 & BVRI & secondary \\
UX Her & 56089.4797 & 0.0003 & BVRI & primary$^{1}$ \\
 & 56093.3568 & 0.0001 & BVRI & secondary$^{1}$ \\
AT Peg & 56146.5608 & 0.0001 & BVRI & primary \\
 & 56153.4397 & 0.0016 & BVRI & primary \\
 & 56157.4471 & 0.0005 & BVRI & secondary \\
 & 55436.5775 & 0.0005 & BR & secondary$^{2}$ \\
 & 55439.4392 & 0.0004 & BR & primary$^{2}$ \\
 & 55442.2968 & 0.0008 & BR & secondary$^{2}$ \\
 & 55447.4616 & 0.0001 & BR & primary$^{2}$ \\
\hline
\end{tabular}
\end{center}
\begin{flushleft}
$^{1}$ \citet{liakos10}, $^{2}$ \citet{liakos14}
\end{flushleft}
\end{table*}
%------------------------------------------------------------End of Table
Two out of the three  systems: 
\astrobj{XZ~Aql} and \astrobj{AT~Peg},
show secular period changes, either period increase or decrease. 
The former may be an indication of mass transfer from the less massive 
to the more massive component or mass loss from the system 
while the latter, mass transfer in the opposite direction. 
We fitted the trends in period changes with parabolae, 
under the assumption of conservative mass transfer between components. 
Furthermore, we analyzed the residuals from the parabolic fit 
for two systems: \astrobj{XZ~Aql} and \astrobj{AT~Peg}. 
Cyclic variations were noticed that could be attributed to the light-time 
effect (LiTE) due to unseen additional components to these systems. 
For \astrobj{UX~Her} we found only 
cyclic variations that may be caused by a companion that is dynamically bound 
to the binary system. The parameters corresponding to companions of the 
three systems were derived using the equations based on the formulation of 
\citet{irwin52, irwin59}. We checked the dynamical stability of the 
configurations 
by using the stability condition given by \citet{harrington77} when an unseen 
third body was assumed to explain the periodic changes in the O-C diagrams.
In each case the orbit of the third bodies has been assumed to be coplanar 
with that of the eclipsing pair.
Errors have been estimated using a specific 
IDL code written by us. The results from the O-C analysis, 
including estimated mass transfer rate are given in Table~\ref{table6}.
The errors were computed by propagating the observational errors 
on the results in the formal manner.
We also give the initial light elements that we used in the 
computation of the orbital phases for each of the systems in this table before 
the results

%-------------------------------------------------- Figure XZ Aql - UX Her O-C

%-------------------------------------------------- Figure AT Peg O-C

%------------------------------------------------------------Table 6
\begin{table*}
\small
\begin{center}
\caption{Results from the O-C analysis.}
\label{table6}
\medskip
\begin{tabular}{lccc}
\hline
\hline
 & \astrobj{XZ~Aql}  & \astrobj{UX~Her} & \astrobj{AT~Peg}\\ 
\hline
T$_{o}$ (HJD) & 2441903.4610 $^{1}$ &  
2439672.3760 $^{2}$ & 2445219.8512 $^{1}$ \\
P (days) & 2.139181 & 1.548853 & 1.1460764 \\ 
\hline
$dM / dt$ [M$_{\odot}$/year] & 6.37(13)x10$^{-8}$ & - & - \\
$dP / dE$ [days/cycle] & 4.33(5)x10$^{-9}$ & - & -1.21(1)x10$^{-9}$ \\
$P_{3}$ [years] & 37.97(96) & 86.81(52) & 33.02(41) \\
$A$ [days] & 0.015(1) & 0.037(2) & 0.016(1) \\
$\omega$ [$^{\circ}$] & 316(18) & 18(1) & 86(11) \\
$e$   & 0.21(10) & 0.41(7) & 0.53(8) \\
$a_{12}sin(i)$ [AU]   & 2.62(150) & 6.99(16) & 2.72(90) \\
$f(m_{3})$ [M$_{\odot}$]   & 0.01(2) & 0.045(3) & 0.02(2) \\
$M_{3,min}$ [M$_{\odot}$]   & 0.52(8) & 0.87(7) & 0.68(5) \\
\hline
\end{tabular}
\end{center}
\begin{flushleft}
$^{1}$ \citet{samus2012}, $^{2}$ \citet{tremko04}
\end{flushleft}
\end{table*}

\subsection{Frequency Analysis}
In order to search for potential pulsations in the data of the studied systems, 
we subtracted the theoretical light curves of the binary model from the 
corresponding observed ones and performed frequency analyses on the 
out-of-eclipse phases of the residuals with the software PERIOD04 v.1.2 
\citep{lenzbreger05} (for further details see \citet{liakos12b}, 
\citet{liakos13}). 
We searched for frequencies up to 80 cycles per day (c/d). 
After the first frequency computation the residuals were subsequently 
pre-whitened for the next one, until the detected frequency had a 
Signal-to-Noise Ratio (S/N) $<$ 4, which is the programme's 
critical trustable limit. The l-degrees of the pulsation modes 
were identified with the software FAMIAS v.1.01 \citep{zima08} 
using theoretical $\delta$ Scuti models 
(MAD - \citet{montalbandupret07}).\\

\begin{figure*}
\centering
\subfloat[]{
  \includegraphics[width=84.00mm]{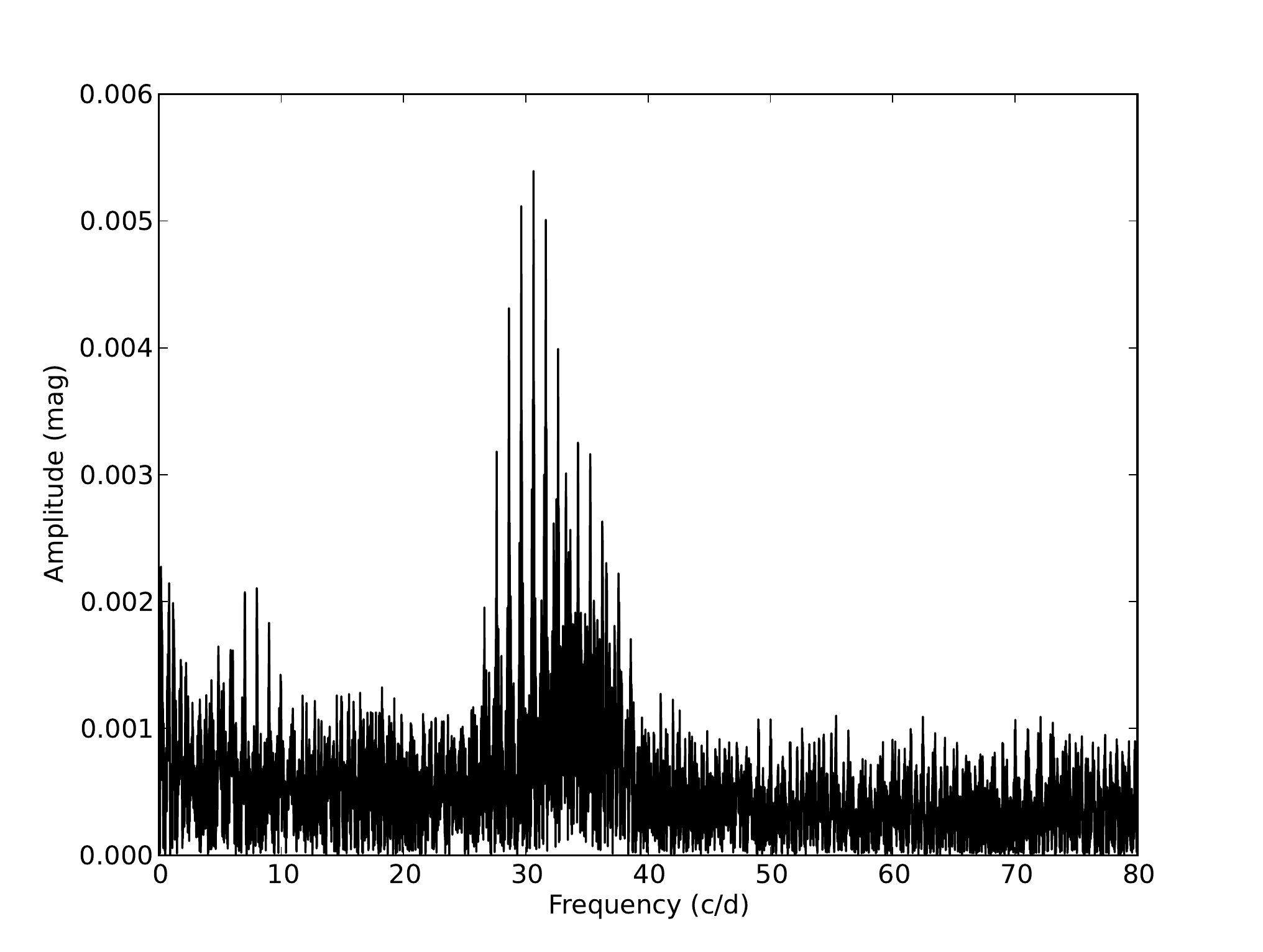}
  \label{fig_pulsfreq}}
\subfloat[]{
  \includegraphics[width=84.00mm]{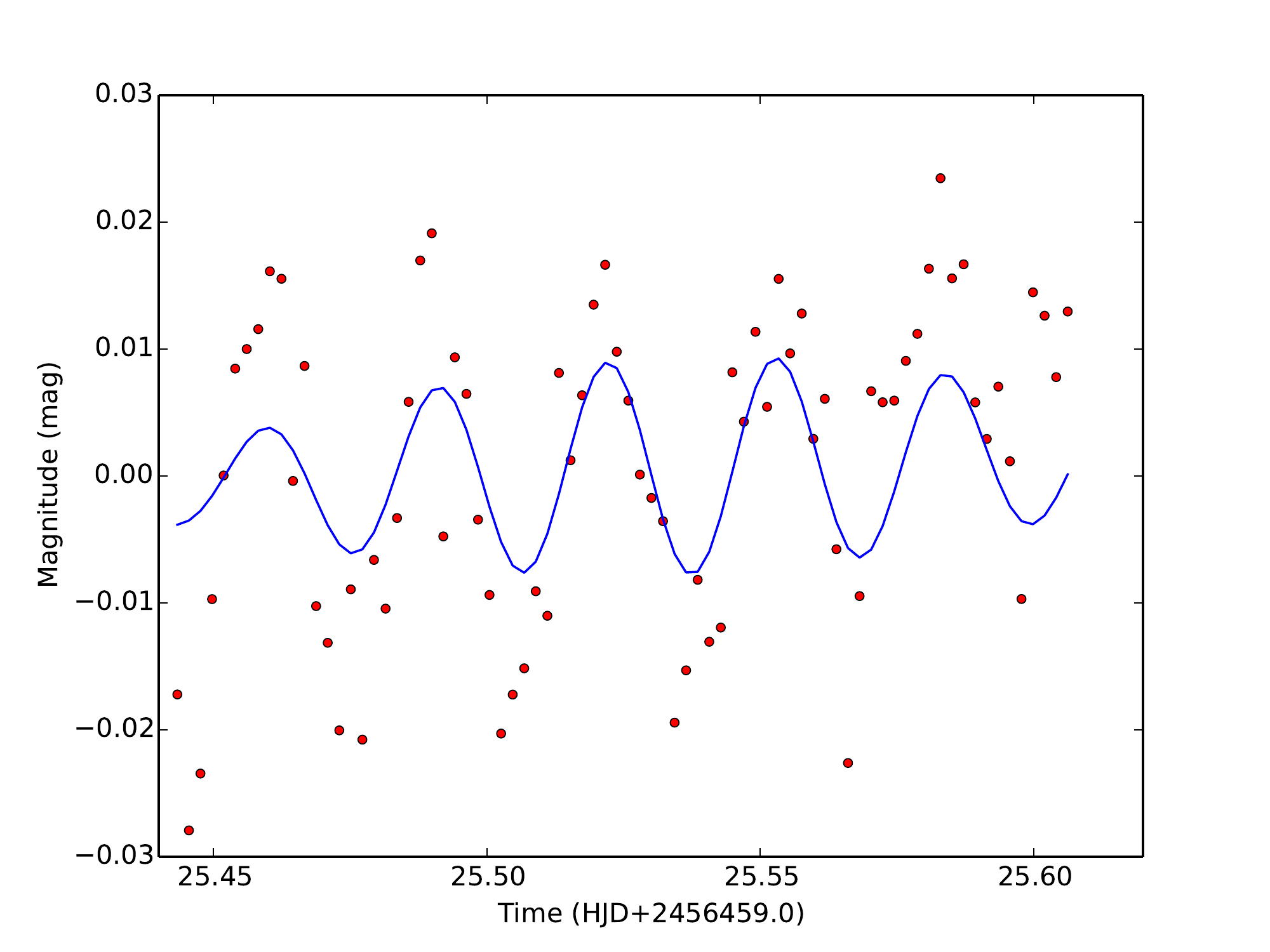}
  \label{fig_pulslc}}
\caption{(a) The periodogram of the $B$ filter data of XZ Aql, 
where the two detected frequencies in the range 30-36 c/d are shown.
(b) Fourier fit on the longest data set for \astrobj{XZ~Aql} 
recorded on the day (JD~2456459) with the $B$ filter.}
\label{fig_pulsation}
\end{figure*}

UX Her and AT Peg did not show any evidence of pulsating behaviour. 
For XZ Aql we found two frequencies in the range 30-36 c/d. 
By taking into account the temperature of the primary component (8770 K), 
the frequency range (3-80 c/d - \citet{breger00}), and 
the spectral types of the
$\delta$ Sct stars (A-F), it can be plausibly suggested that this component 
is a $\delta$ Sct type pulsator. The pulsation signal is present in 
all filters, and its amplitude decreases from $B$ to $I$.
Frequency analysis includes $\sim$1300 points per filter coming from 
19 nights of observations in a time span of 73 days. The data of the first 
17 nights were obtained in a time span of 28 days, sufficient 
to detect quick-pulsation modes. Finally, given that the system is a 
conventional semi-detached system (i.e. the more massive component, 
the pulsator in this case, is the mass gainer), 
it can be also categorized as a typical oEA system 
\citep{mkrtichian02}. Furthermore, two more frequencies were found 
(0.50 and 0.93 c/d) but they were also detected with approximately 
the same values in the comparison-check light curves. 
So, we conclude that their origin is not connected with the true pulsations, and 
they can be considered as artifacts (e.g. observational drift). 
Table~\ref{table7} includes the frequency values (f), the amplitudes (A), 
the phases ($\Phi$), the S/N, and the l-degrees. 
In Fig.~\ref{fig_pulsation} the periodogram, and  
the Fourier fit on the data of an individual night are presented. 
Although  the observed amplitude is rather low to reach a firm conclusion, 
there are more than one freqency with sufficient S/N in the range 
where $\delta$ Scuti stars pulsate and the beating seems to be obvious. 
More precise future observations will provide further evidence 
for the existence of heat-driven pulsations in this system.

%------------------------------------------------------------Table 7
\begin{table*}
\small
\begin{center}
\caption{Results from the frequency analysis on \astrobj{XZ~Aql} data.}
\label{table7}
\medskip
\begin{tabular}{cccccc}
\hline
\hline
 l-degree & Filter & f [c/d] & A [mmag] & $\Phi$ [2 $\pi$ rad] & S/N\\ 
\hline
\multirow{4}{*}{3} & $B$ & 30.631(1) & 5.4(4) & 0.51(1) & 7.8 \\
 & V & 30.633(1) & 4.4(4) & 0.46(1) & 8.1 \\
 & R & 30.632(1) & 3.2(4) & 0.48(2) & 5.3 \\
 & I & 30.635(1) & 2.6(3) & 0.44(2) & 7.9 \\
\hline
\multirow{4}{*}{0} & $B$ & 35.247(1) & 3.2(4) & 0.64(2) & 4.0 \\
 & V & 35.250(1) & 2.4(3) & 0.57(2) & 4.2 \\
 & R & 35.289(1) & 1.9(4) & 0.73(3) & 4.3 \\
 & I & 35.240(1) & 1.9(3) & 0.81(2) & 3.9 \\
\hline
\end{tabular}
\end{center}
\end{table*}
%--------------------------------------------------------------End of Table-7

%-----------------------------------------------Figure - Frequency Analysis

\subsection{Evolutionary Status}
\begin{figure*}
\center
\includegraphics[width=150.00mm]{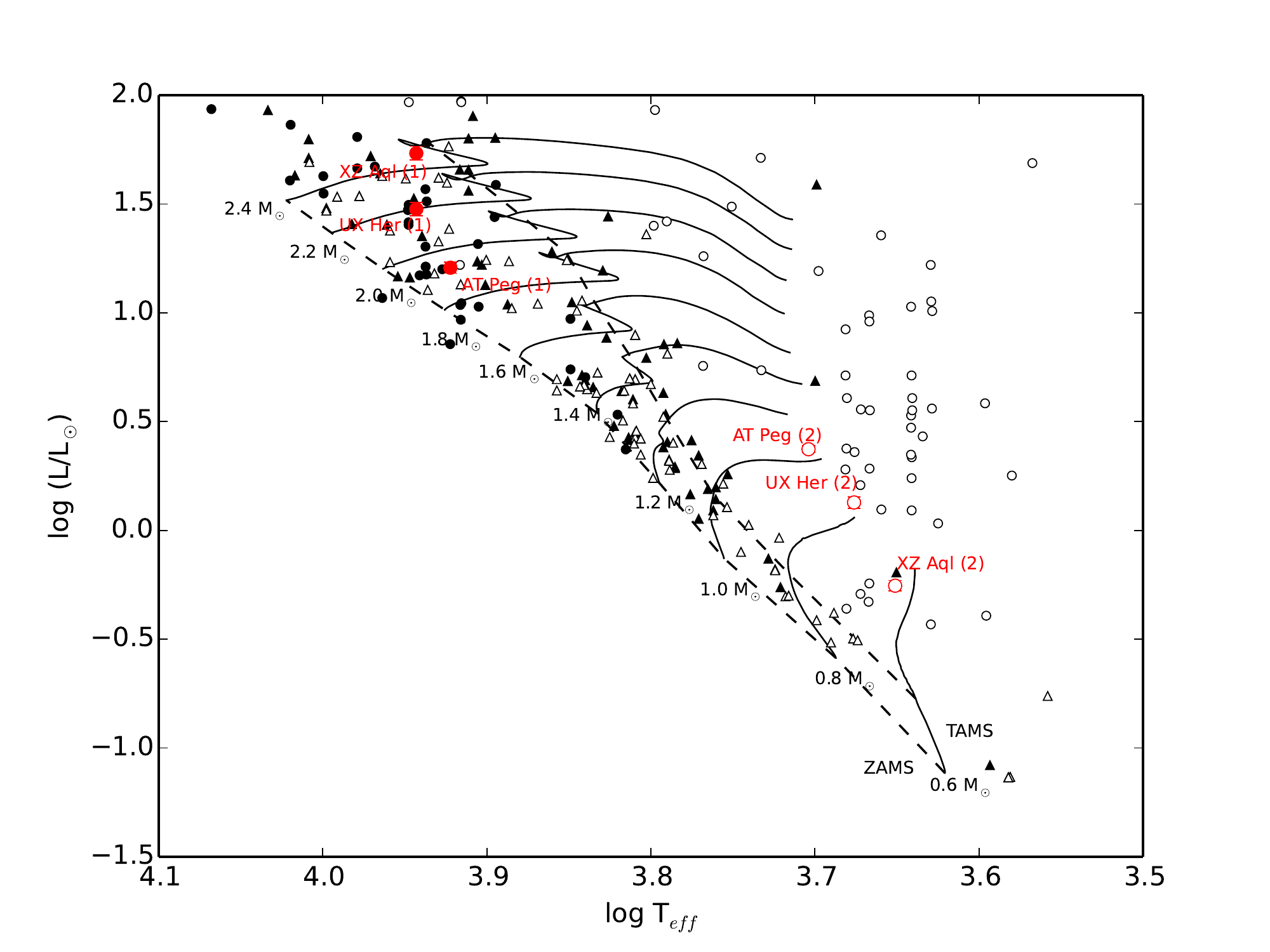}
\caption{The positions of detached (primaries in filled, 
secondaries in unfilled triangles) and semi-detached (primaries in filled, 
secondaries in unfilled circles)  binaries in the HRD. Parameters of 
detached systems taken from \citet{torres2010}, while these for 
semi-detached  ones from \citet{surkova2004}.Parametres of stars analyzed in 
this work  are marked with larger symbols (red colored in the electronic 
version). The evolutionary tracks for a dense grid of masses, 
done by interpolation between the existing tracks of the Geneva stellar models 
for the mass range 0.5-3.5 M$_{\circ}$ \citep{mowlavi2012}, 
are plotted with lines. ZAMS and TAMS are also indicated with dashed lines. 
The location of the components of \astrobj{AT Peg} is based on the 
absolute measurements from the WD fits using our own radial 
velocity observations.}
\label{fig_HRD}
\end{figure*}

\begin{figure*}
\centering
\subfloat[]{
  \centering
  \includegraphics[width=84.00mm]{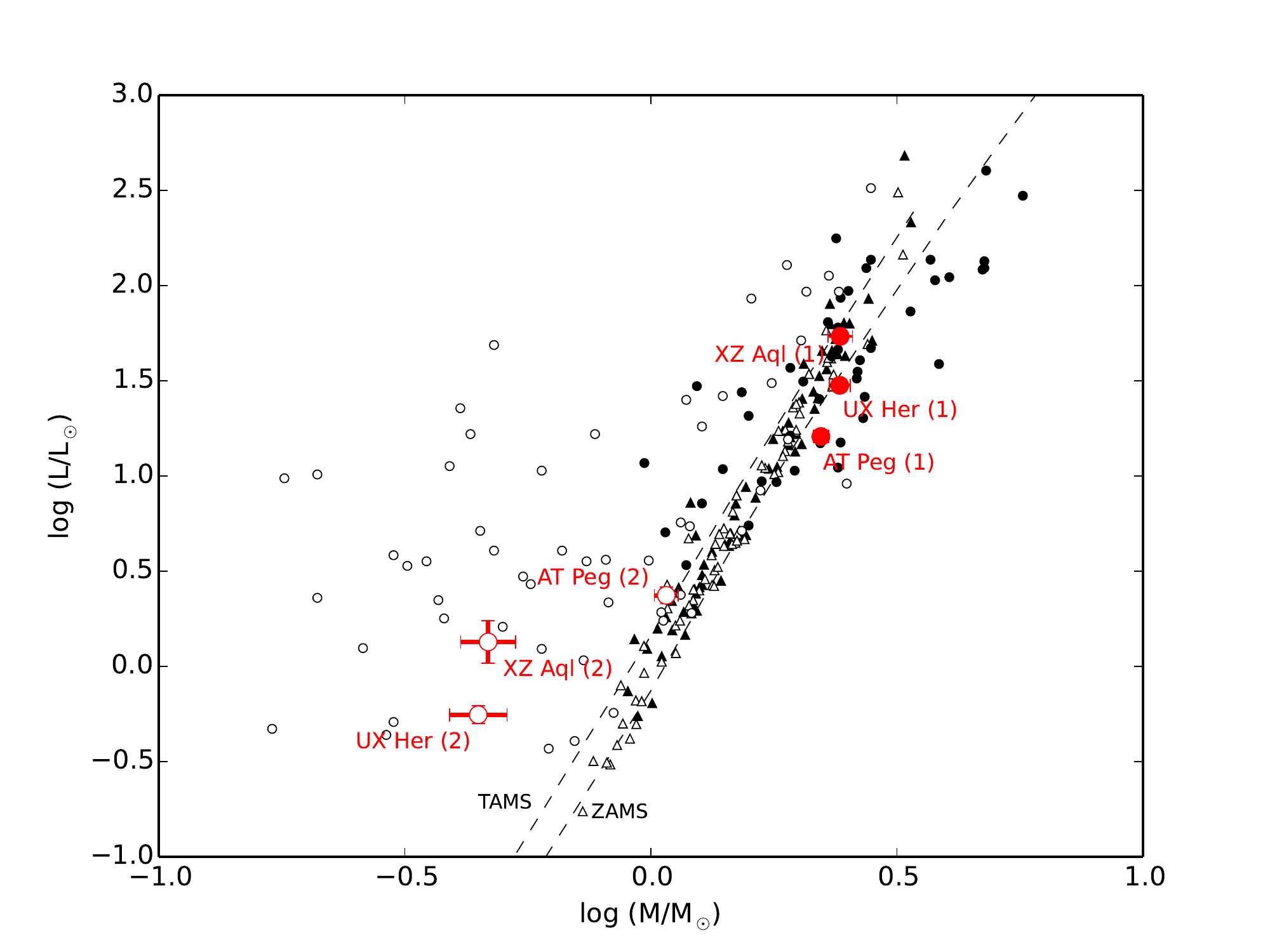}
  \label{fig_MLD}}
\subfloat[]{
  \centering
  \includegraphics[width=84.00mm]{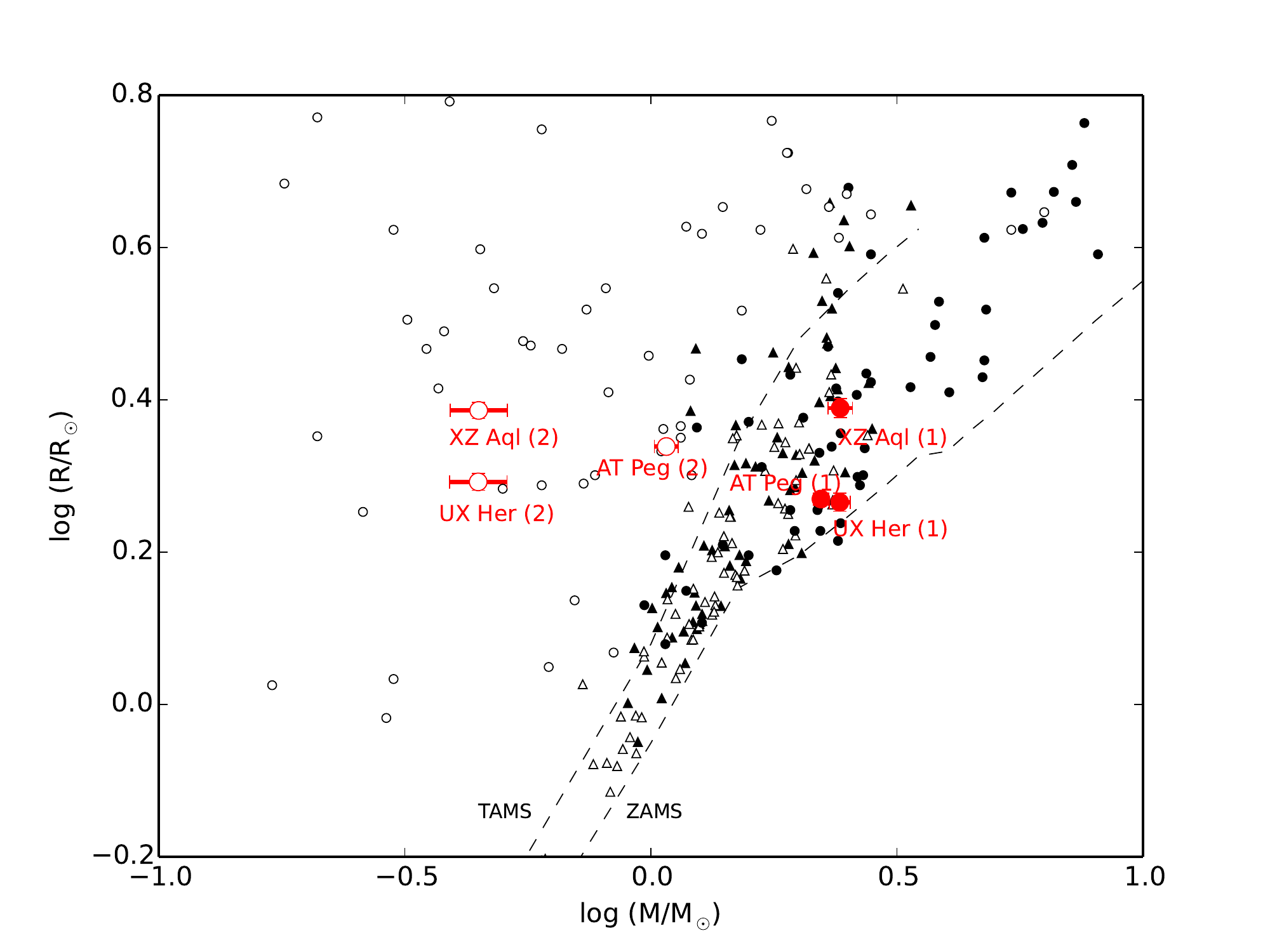}
  \label{fig_MRD}}
\caption{Mass-Luminosity (a) and Mass-Radius (b) diagrams. 
ZAMS and TAMS computed from the theoretical models by \citet{ekstrom2012} 
are plotted by dashed lines.  Positions of components in detached, 
semi-detached and these in systems derived in this work are also shown. 
Symbols meaning is the same as in Fig. \ref{fig_HRD}.} 
\label{fig_MLD_MRD}
\end{figure*}

We used the Geneva stellar models \citep{ekstrom2012} to investigate the 
evolutionary states of the components of all the studied systems. 
We generated a grid of evolutionary tracks for a suitable 
range of fixed masses spanning the measured values, 
by making use of interactive  tools provided by a web 
interface\footnote{http://obswww.unige.ch/Recherche/evoldb/index/}. We 
fixed metallicity at solar composition (Z = 0.014) and generated non-rotating 
models. In order to populate the Hertzsprung-Russell diagram (HRD) 
with observed stars in detached, semi-detached and near-contact eclipsing 
binaries, we collected the parameters of components in detached systems 
from \citet{torres2010} and those in semi-detached binaries from 
\citet{surkova2004}. From both catalogues we have selected only the binaries 
with spectroscopically determined mass ratios. In Fig.~\ref{fig_HRD} we show 
the computed Geneva evolutionary tracks, parameters of 
components of selected detached binaries (primaries are indicated with filled 
while secondaries with unfilled triangles), semi-detached and 
near-contact binaries (primaries are indicated with filled, 
secondaries with unfilled circles) and our program stars 
(indicated with marks selected according to their nature and component 
type and in red color in the electronic version) on the HRD. 
We also present the Mass-Luminosity (MLD) and Mass-Radius (MRD) diagrams 
in Fig.~\ref{fig_MLD_MRD} following the same symbol types as these used in 
Fig.~\ref{fig_HRD}. It should be borne in mind that the stellar models 
are valid only for single stars, or stars in binaries which do not interact 
strongly, while the stars in our sample can not be assumed to be free from 
stellar interactions in a binary system. The implications of this fact are 
clear when the positions of the evolved secondary components in these 
systems are compared with the evolutionary tracks given for their masses 
because they transfer material to their more massive and rather 
unevolved counterparts. The lines indicating the positions of the 
Zero Age Main Sequence (ZAMS) and the Terminal Age Main Sequence 
(TAMS) are shown in these figures. 
ZAMS is defined as the time when the hydrogen mass fraction 
in the center (X$_{c}$) has decreased by 0.25\% for a newborn star 
with the solar composition while TAMS is set at the time when X$_{c}$ will be 
 of the value 10$^{-5}$ for a star starting its life with solar composition. 
Following the above definition, a star will be on the ZAMS when 
thermal equilibrium is achieved and the star has burnt 0.25\% of its 
hyrdogen in the core \citep{ekstrom2012,mowlavi2012}. 

%------------------------------------------------------------- Figure HRD

%------------------------------------------------------------- Figure MLD-MRD

\section{Results \& Discussion}
In this study we have performed thorough analyses of the light curves 
and period variations of three close binary systems. Analysis of our 
light curves was based on combined photometric and spectroscopic data. 
The mass ratios of the studied systems were determined from our own 
spectroscopic observations. We made use of the Wilson-Devinney 
algorithm to derive the best fits to new, multicolor 
light curves, and based on obtained results we calculated the absolute 
parameters of components. 
The complementary information about the systems has been derived by 
analyzing their period behaviour.

\subsection{\astrobj{XZ~Aql}}
The light curve of {\astrobj{XZ~Aql} has equal levels of light maxima, 
and we were able to obtain 
its solution that required  neither a spot nora  third light. 
This is a high inclination (i=86 deg), semi-detached binary with the less 
massive secondary filling its Roche lobe, while the primary component 
is well inside its Roche lobe. The frequency analysis performed after 
having removed the binarity effects from its light curves hints that 
this A-type primary is a $\delta$ Sct type pulsator. 
Due to a significant temperature difference 
between components, the contribution of the secondary star to the total system 
light is small: only 2\% in the $B$ filter and  
about 13\% in the $I$ filter. \\

From the O-C analysis we found a parabolic change in the eclipse 
timings combined 
with a periodic variation that could be due to an unseen third body as 
previously suggested by \citet{soydugan06}. We made trial computations 
with a third light, however, its resulting intensity was negligible, 
reaching only 1\% in the $I$ filter. Assuming that this third body is in 
coplanar orbit with the binary and that it is an MS star, 
its luminosity contribution to the total light would only be 0.2\%.
We found that the orbit of this third companion would be stable according to 
Harrington's criterion \citep{harrington77}.
We have found a parabolic relation in the residuals that may be due to 
mass transfer from the less massive component to the more massive one. 
When the corresponding orbital period changes are removed, 
one might argue that there is a possibility 
of a second periodicity in the O-C points, but the number of data points is 
not sufficient to prove it. Alternatively, the periodic variation of the 
orbital period could also be explained with the quadrupole moment 
variation of the secondary.\\ 

Our results from the orbital period variation analysis also
support the finding that the secondary component fills its 
Roche lobe and transfers mass to the more massive primary conservatively. 
The positions of its component on the HRD, MLD, and MRD based on the absolute 
parameters computed for each of the components as a result of the light curve
modeling support this finding. Both the computed masses and 
the effective temperatures from our fit are qualitatively consistent with 
evolutionary tracks, 
which also point to a main-sequence primary half way between the
ZAMS and the TAMS with $\sim$ 2.42 M$_{\odot}$, and an evolved red giant 
secondary, somewhat more massive ($\sim$ 0.65 M$_{\odot}$) 
than our computed 
value ($\sim$ 0.45 M$_{\odot}$). This discrepancy can be explained by
the mass transfer from the secondary to the primary star, 
the time scale of which should be
less than 10 Myr assuming a constant rate for the mass transfer. The fact that
the main sequence primary has not gained all the mass the secondary transfers
can only be explained by a non-conservative mass loss.

\subsection{\astrobj{UX~Her}}
Our light curve modelling for \astrobj{UX~Her} used the new value 
for the spectroscopic mass ratio resulting from our DAO data  q$_{sp}$ = 0.184. 
It is significantly smaller than the previous determination 
from photometry alone as a result of the q-search by \citep{gojko06}. 
We obtained 
a semi-detached configuration for this system with the less massive component 
filling its Roche lobe while the primary being well within it. 
Therefore, mainly due to the lower q value, the derived secondary mass is also 
smaller than that previously found \citep{lazaro97,gojko06}, both considered 
this system as a detached one based on either photometrically or 
spectroscopically determined mass ratios from photographic plates. 
The evolutionary state of \astrobj{UX~Her}  components deduced 
from their positions on the HRD, MLD and MRD diagrams point to the 
interpretation that the primary star is still on the main sequence, 
while the secondary is the more evolved star.\\

We did not find a significant quadratic component to the period variation, 
therefore we conclude there is no evidence of mass transfer.
The cyclic period variations can more plausibly be explained by a third body 
of mass M$_{3}$ = 0.87 M$_{\odot}$, orbiting the center of mass on a significantly
 eccentric orbit (e = 0.41). \citet{tremko04} found a smaller 
mass, less than half of the value that we found (0.30 M$_{\odot}$).
As a test, we also performed computations 
adding a third light as a free parameter. It turned out that contribution 
of this hypothetical tertiary to the total light is negligible: 
it reached about 1\% only in the $I$ filter so therefore we present 
the solution 
without l$_3$ as the final one. If we assume that this body has a coplanar
orbit with that of the eclipsing binary and that it is an M5 star, its 
luminosity contribution will be less than 1\% (0.75\% according to our 
calculations), which is below our detection limits with the photometry.
Previously \citet{selam07} also found cyclic changes in the orbital period 
in \astrobj{UX~Her}. They attributed the variation to magnetic activity, 
which would also 
explain the observed asymmetries in the light curve. However, the 
latest CCD data have a very low scatter around our best fit. 
More scatter would be expected in the case of magnetic activity because it
would complicate the measurement of the eclipse times from asymmetric
minimum profiles due to the surface spots therefore causing more errors. Further
observations of the system covering a longer time base will be needed 
to make sure that the period and the amplitude of the variation change from 
one cycle to another, which would be the case in the presence of 
strong magnetic activity. Otherwise, the variation could be argued to cause 
from the gravitational pull of an unseen third body. 
In additon, the temperature of the primary is too high (8770 K) 
to expect cool surface spots. Though it would be somewhat reasonable to 
locate spots on the cooler secondary, 
our non-spotted solution satisfactorily fits the 
observations. Therefore, we give here only the third body solution 
because we do not have further evidence (e.g. variations in magnetic activity 
indicators) supporting the magnetic activity argument. Moreover, the dynamical 
stability test according to Harrington's criterion \citep{harrington77} points 
to a stable orbit for the suggested third body, thus supporting the second 
hypothesis.

\subsection{\astrobj{AT~Peg}}
Light curve modelling for \astrobj{AT~Peg} proceeded without difficulty 
for both radial velocity data sets (our own and that of 
\citet{maxtedetal94}).
A convergence was quickly found for a model that required neither 
a third light nor a spot. Theoretical light curves fit observed ones very well 
with just very small discrepancies visible around the primary minimum 
but only in $B$ and $I$ filters. We arrived at the semi-detached solution with 
the less massive secondary filling the Roche lobe and the primary well 
within it. Absolute parameters and their errors differ only within 
a few percent between the two models obtained by using two different 
radial velocity data sets. \\

The O-C analysis shows a parabolic trend. A 
periodic relation for the residuals can also be asserted.
The quadratic term cannot be interpreted as being 
due to conservative mass transfer between components as it would require the 
more massive star to be the mass loser, in contradiction with the results 
from light curve modeling. This discrepancy can be explained by 
non-conservative mass loss from the system or losing of angular momentum 
via magnetic breaking. Such systems with orbital period decrease 
may be the progenitors of contact binaries \citep{bradstreet94,qian2000}, 
because the fill-out factor of the primary may increase 
while the orbital period decreases, eventually causing the primary 
to fill its Roche lobe. We do not have evidence for strong 
magnetic activity such as light curve modulations and asymmetries 
in the light curves due to stellar spots. 
The spectral window of our spectral observations is also limited to 
5000-5250 \AA ~region, where we have not observed a signature of activity or 
stellar wind in the resolution our spectra were taken. 
The spectral type of the primary (A4) would not support this argument either, 
assuming it is a normal star without any kind of pecularity. The positions
of the components relative to the evolutionary tracks on the HRD indicate 
that the primary might have lost some mass. The 
evolutionary model value with no mass transfer is $\sim$ 1.90 M$_{\odot}$ for 
the primary component, while the computed mass from the light curve analysis 
is 2.22 M$_{\odot}$ and the mass of the secondary is consistent in both models 
($\sim$ 1.08 M$_{\odot}$).\\

The residuals from the quadratic fit may be alleged to follow 
a periodic behavior, which can be attributed to an unseen body 
(M$_{3,min}$ = 0.68 M$_{\odot}$) gravitionally bound to the system, 
orbiting its center of mass once in every $\sim$33 years on a 
a significantly eccentric (e = 0.53) orbit. The relatively low mass of 
the tertiary can explain its negligible contribution to the total light. 
Additional bodies have been plausibly argued to extract momentum 
from the binary, causing the orbit to shrink and hence the orbital period 
to decrease \citep{yangwei09}. This could be at least a part of the 
reason that causes a period decrease although the direction of the 
mass transfer is towards the more massive primary.


%% Included in this acknowledgments section are examples of the
%% AASTeX hypertext markup commands. Use \url without the optional [HREF]
%% argument when you want to print the url directly in the text. Otherwise,
%% use either \url or \anchor, with the HREF as the first argument and the
%% text to be printed in the second.

\acknowledgments
\section{Acknowledgments}
This work was partially supported by the NCN grant No. 2012/07/B/ST9/04432.
This work was performed in the framework of PROTEAS project within GSRT’s 
KRIPIS action for A.L., funded by Greece and the European Regional Development 
Fund of the European Union under the O.P. Competitiveness and Entrepreneurship, 
NSRF 2007-2013 and the Regional Operational Program of Attica. 
We gratefully acknowledge invaluable discussions with Prof. Selim O. Selam 
and his suggestions. We thank Prof. Slavek Rucinski for his comments on the 
broadening function. We also thank all the observers participating in the 
observations of our targets, and the staff members at the DAO and UoA.\\




\end{document}